\documentclass[cup6b]{cupbook}  

\input psfig
\def\fig #1, #2, #3 {
\smallskip
\centerline{\psfig{figure=#1,height=#2 in,width=#3 in}}
}

\def\d{{\rm d}}
\newcommand{\iras}{{\sl IRAS\/}}
\newcommand{\etal}{{\sl et al.\/}}
\newcommand{\mpc}{\ifmmode{\,h^{-1}\,\mathrm {Mpc}} \else
{$\,h^{-1}\,$Mpc}\fi}
\newcommand{\kms}{\ifmmode {{\rm \ km \ s^{-1}}}
\else{$\mathrm {km \ s^{-1}}$}\fi}
\newcommand{\bfv}{{\bf v}}
\newcommand{\bfr}{{\bf r}}
\newcommand{\bfx}{{\bf x}}
\newcommand{\bfk}{{\bf k}}
\newcommand{\bfs}{{\bf s}}
\newcommand{\vev}[1]{\left\langle #1 \right\rangle}
\newcommand{\gtsima}{$\; \buildrel > \over \sim \;$}
\newcommand{\ltsima}{$\; \buildrel < \over \sim \;$}
\newcommand{\simgt}{\lower.5ex\hbox{\gtsima}}
\newcommand{\simlt}{\lower.5ex\hbox{\ltsima}}
\def\pmb#1{\setbox0=\hbox{#1}%
\kern-.025em\copy0\kern-\wd0
\kern.05em\copy0\kern-\wd0
\kern-.025em\raise.0433em\box0}
\begin{document}

\author[M. A. Strauss]{Michael A. STRAUSS\footnote{Alfred P. Sloan
Foundation Fellow}\\Princeton University Observatory}
\chapter{Recent Advances in Redshift Surveys of the Local Universe} 

\begin{abstract}
I review progress in the past few years in studying the large-scale
structure of the universe through redshift surveys of galaxies.  Of
the many statistical methods used to describe the galaxy distribution,
I concentrate here on the power spectrum, and go into some detail
about the factors which complicate (and make interesting!) its
interpretation, such as redshift space distortions, non-linear effects,
and the relative bias of galaxies and dark matter.  I also discuss two
large redshift surveys which are just starting, the Sloan Digital Sky
Survey, and the Two Degree Field Redshift Survey, which promise to
increase the number of redshifts measured of galaxies in uniform
surveys by more than an order of magnitude.
\end{abstract}

\section{Introduction}
\label{sec:intro}
The past two decades have seen an explosion of our knowledge in many areas
of observational cosmology.  One of the most significant has been
increases in our understanding of the distribution of galaxies in the
nearby universe (defined in the context of the present review as $z
\simlt 0.1$, where cosmological and evolutionary corrections can be
neglected, for the most part).  In addition to allowing us to do {\it
cosmography}, whereby the structures and forms the galaxies find
themselves in are mapped and catalogued, the several orders of
magnitude increase in the number of galaxies with measured redshifts
over this time period allows us to do quantitative {\it cosmology},
whereby we put specific constraints on models for structure formation
and the various parameters which are input to the
Friedman-Robertson-Walker metric.  In addition to the much larger
number of measured redshifts in complete samples of galaxies available
now, there have been great advances in our theoretical understanding
of the various statistics that have been measured from redshift
surveys. 

  We shall not attempt in this review to give a complete and thorough
summary of the entire subject of what can be learned from redshift
surveys.  Jeff Willick and I reviewed the field in a comprehensive
article which covers material through the end of 1994 (Strauss \&
Willick 1995, hereafter SW), and I will try to avoid duplicating too
much of the discussion found therein.  Other recent reviews that
discuss redshift surveys include those of Giovanelli \& Haynes (1991),
Dekel (1994), Borgani (1995), Efstathiou (1995, 1996), 
and Guzzo (1996).
The emphasis in this review will be on developments which have
occurred since the writing of SW, with special emphasis on future
redshift surveys and what they will be able to measure.

\section{Varieties of Redshift Surveys}
\label{sec:variety}

In order to be useful for any sort of statistical work, a redshift
survey must have a well-defined selection function in the most general
sense of the term.  That is, the selection criteria of the galaxies
whose redshifts are included must be objective and quantifiable. This
is not the same as saying that the redshift survey must be complete in
some sense, just that its incompleteness follows some known rule(s),
such that the fraction of galaxies with redshifts is known, at least
statistically, as a function of the observational properties of a
galaxy.  In practice, this means that the following must be known for
any redshift survey, and the galaxies contained therein, if it is to
be useful:
\begin{itemize}
\item The region of sky covered by the survey.
\item The observational criterion or criteria by which galaxies have
been selected for redshifts, such as apparent magnitude in a given band,
diameter, surface brightness cuts, emission-line strength\footnote{relevant,
e.g., in objective prism surveys for emission-line galaxies or
quasars.}, etc.  
\item The value of the relevant quantity or quantities for
all objects in the sample. 
\item Limits in the above quantity or quantities.  This may very well
be a function of position on the sky, such as in the case of an
apparent magnitude-limited survey in optical bands, in the presence of
Galactic extinction.  These limits determine the depth of the survey,
which can be quantified in terms of the expected number distribution
of galaxies as a function of redshift. 
\item The fraction of galaxies for which redshifts are actually
obtained.  Again, this could be a function of position on the sky, or
could be a fixed fraction over the survey area.  If the fraction is
close to 100\%, we call this a complete redshift survey. 
\item The positions and redshifts of the individual galaxies. 
\end{itemize}

Speaking loosely, then, a survey is characterized by its solid angle
coverage, its depth, its sampling rate, and the method of selection of
the galaxies.  Comprehensive lists of redshift surveys are given in the
various reviews listed in the Introduction, including SW; we will not
repeat this here.  However, we briefly summarize some of the redshift
surveys we will find ourselves referring to through this review, with
apologies to those whose surveys we do not have space to discuss
here. 

  There have been a few surveys which have covered close to the entire
celestial sphere; as we will see, these have been very important both
for cosmography, and for making dynamical predictions of the peculiar
velocity field in the nearby universe.  Early work in this direction
was done by Yahil, Sandage, \& Tammann (1980) with the Revised
Shapley-Ames catalog of galaxies (cf., Sandage \& Tammann 1981), but
this sample only extended to redshifts of 4000 \kms, and was affected
strongly by the Galactic zone of avoidance.  The {\it Infrared
Astronomical Satellite}, or \iras, flew in 1983, and scanned the
entire sky at $\sim 1'$ resolution in four broad bands centered at 12,
25, 60, and 100$\mu$m (cf., \iras\ Point Source Catalog Explanatory
Supplement 1988 for details). An ordinary spiral galaxy with a
moderate amount of star formation emits strongly at 60$\mu$m as
thermal emission from dust heated by the interstellar radiation field.
Because of the all-sky nature of the \iras\ survey, and the
transparency of the dust of the Milky Way to 60$\mu$m radiation, a
galaxy sample selected at 60$\mu$m has uniform and full sky coverage.
Two groups have carried out extensive full-sky redshift surveys based
on these data: one centered in Berkeley, doing redshift surveys
complete at 60$\mu$m to 1.936 Jy and subsequently to 1.2 Jy (Strauss
\etal\ 1990, 1992c; Fisher \etal\ 1995a), and the other a mostly
British collaboration which obtained redshifts of 1 in 6 galaxies to a
flux limit of 0.6 Jy (Rowan-Robinson \etal\ 1990; Lawrence \etal\
1996).  This latter collaboration, referred to as QDOT for the
initials of the institutions of the investigators, is currently
extending their effort to measure redshifts for a {\it complete}
sample of galaxies to 0.6 Jy (15,500 galaxies) at 60$\mu$m. See {\tt
http://www-astro.physics.ox.ac.uk/$\sim$wjs/pscz.html\/} for the
latest details.  In practice, the \iras\ surveys are limited by the
effects of Galactic extinction at very low latitudes (if you can't see
the galaxy optically, you certainly cannot measure a redshift!) and
systematic effects in the \iras\ Point Source Catalog in regions of
very high source density (mainly confusion and hysteresis); these
surveys thus cover between 80 and 90\% of the sky.

  Elliptical galaxies are very faint for the most part in the infrared
bands, and are therefore essentially absent from the \iras\
surveys.  Given that the cores of clusters of galaxies are dominated by
elliptical galaxies (e.g., Dressler 1980, 1984; Postman \& Geller
1984; Whitmore, Gilmore, \& Jones 1993), the number density of galaxies in the 
cores of clusters is systematically underestimated in \iras, as is in
fact borne out in direct comparisons of \iras\ and optically selected
samples (cf., Strauss \etal\ 1992a; Santiago \& Strauss 1992).  This,
together with the rather sparse sampling of the galaxy distribution by
the \iras\ satellite, motivated the compilation of the {\it Optical
Redshift Survey} (ORS; Santiago \etal\ 1995, 1996; Hermit \etal\
1996), which selects galaxies from the Uppsala Galaxy Catalogue
(Nilson 1973), the ESO Galaxy Catalogue (Lauberts 1982) and its
photometric counterpart (Lauberts \& Valentijn 1989) and the Extension
to the Southern Galaxy Catalogue (Corwin \& Skiff 1996).  Galactic
extinction restricts the survey to Galactic latitudes $|b| >
20^\circ$.  The depth is comparable to that of the original CfA survey
(Davis \etal\ 1980; Huchra \etal\ 1983) but with 4.5 times the sky
coverage.  Santiago \etal\ (1996) detail the effort required to tie the
three parts of the survey together, and the corrections made for
Galactic extinction. 

\begin{figure}
\centerline{\psfig{file=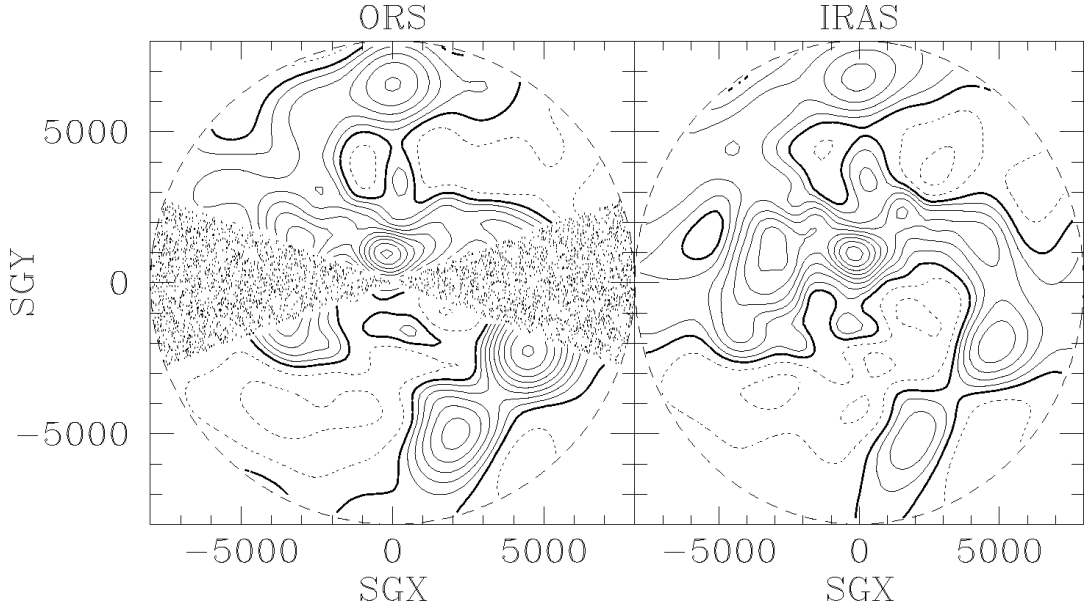,width=7cm,angle=270,bbllx=47.8pt,bblly=270.6pt,bburx=552pt,bbury=627.2pt}}
\caption{Galaxy isodensity contours in the Supergalactic Plane for the
ORS (left) and \iras\ 1.2 Jy redshift surveys (right).  The local
group is at the center of each map. The smoothing in the two cases is
the same, and increases with distance from the center, therefore
relative strength of features at different distances from the center
can be misleading.  The heavy contour is at the mean density; dotted
contours are underdense relative to the mean. The zone of avoidance is
indicated in the case of the ORS sample.}
\label{fig:sgp}
\end{figure}

  The Local Group is a member of the Local Supercluster, a highly
flattened structure which extends at least to 3000 \kms\ from us.
Indeed, many of the other dramatic superclusters in the nearby
universe are found in, or at least are intersected by, the plane
defined by the Local Supercluster, including the Perseus-Pisces
Supercluster, the Hydra-Centaurus and Pavo-Indus-Telescopium
Superclusters (almost certainly a single structure bisected by the
zone of avoidance) and the Coma-A$\,$1367 Supercluster.
Fig.~\ref{fig:sgp} shows isodensity contours of the galaxy
distribution in the Supergalactic plane, in the \iras\ 1.2 Jy (left)
and ORS (right).  The smoothing is Gaussian with $\sigma = 400 \kms$
at low redshift, increasing like the \iras\ mean interparticle spacing
at greater distances (thus the smoothing is the same in the two
surveys).  Mean density is indicated with a heavy line; contours above
mean density are spaced logarithmically, with every third contour
representing a factor of two in density.  Dotted contours are at 0.66
and 0.33 of the mean density.  Fingers of God associated with rich
clusters have been collapsed to a common redshift; otherwise the
galaxies are placed at the distances indicated by their redshifts in
the Local Group frame (i.e., no correction for peculiar velocities
have been made).  The low latitude regions of the ORS map are masked
out, because of the absence of galaxies there.  It is perhaps not
surprising that the large-scale distribution of galaxies in these two
surveys is similar; there is of course a great deal of overlap in the
two samples, and of course the majority of galaxies in an optically
magnitude limited sample are spirals. 

  The smoothing here is not uniform, but increases with distance from
the origin as the samples become sparser.  Therefore, the relative
strength of features seen at different distances from the origin can
be misleading.  We could avoid this by choosing a single large
smoothing scale.  Alternatively, one can smooth with a
noise-suppressing filter; a Wiener filter (cf., Hoffman, these
proceedings) still gives an effective smoothing that increases with
distance from the origin, but a simple variant on that, called the
power-preserving filter (cf., Yahil, these proceedings) gives a
constant smoothing with distance. See SW for maps of the \iras\
density field with this smoothing scheme. 

  There are two major recent surveys which go somewhat deeper than the
surveys discussed above, although they are far from full-sky surveys.
The CfA2 survey (cf., Geller \& Huchra 1988; 1989 for early reviews)
covers 2.95 ster in the Northern Hemisphere, and includes all galaxies
from the Zwicky \etal\ (1961-1968) catalog with $m_Z \le 15.5$.  A
parallel effort (da Costa \etal\ 1994a) in the Southern Hemisphere is
covering 1.13 ster to the same depth of the CfA2 survey, based on
scans of sky survey plates.  The two surveys have been analyzed
together to determine the power spectrum of galaxies (da Costa \etal\
1994b) and the small-scale velocity dispersion of galaxies (Marzke
\etal\ 1995).

On smaller angular scales, but going deeper, are several surveys.
Shectman \etal\ (1996) have used a multi-object spectrograph on the
Las Campanas 2.5m Du Pont Telescope to obtain redshifts for 26,418
galaxies to $r \approx 17.5$ (the Las Campanas Redshift Survey, or
LCRS).  The photometry is based on CCD drift-scan data obtained by the
same workers on the Swope 40'' telescope at Las Campanas. This sample
is not complete: fields in which to do spectroscopy were laid down on
a regular grid in a series of six $\sim 90^\circ \times 1.5^\circ$
wide stripes across the sky, and the sampling rate was simply the
ratio of the number of galaxies available to their magnitude limit, to
the number of fibers of the spectrograph, averaging roughly 70\% over
their fields. This survey, covering a total of $\sim 700$ square
degrees, has roughly a factor of two more redshifts than any other
single redshift survey of galaxies.

\begin{figure}
\fig 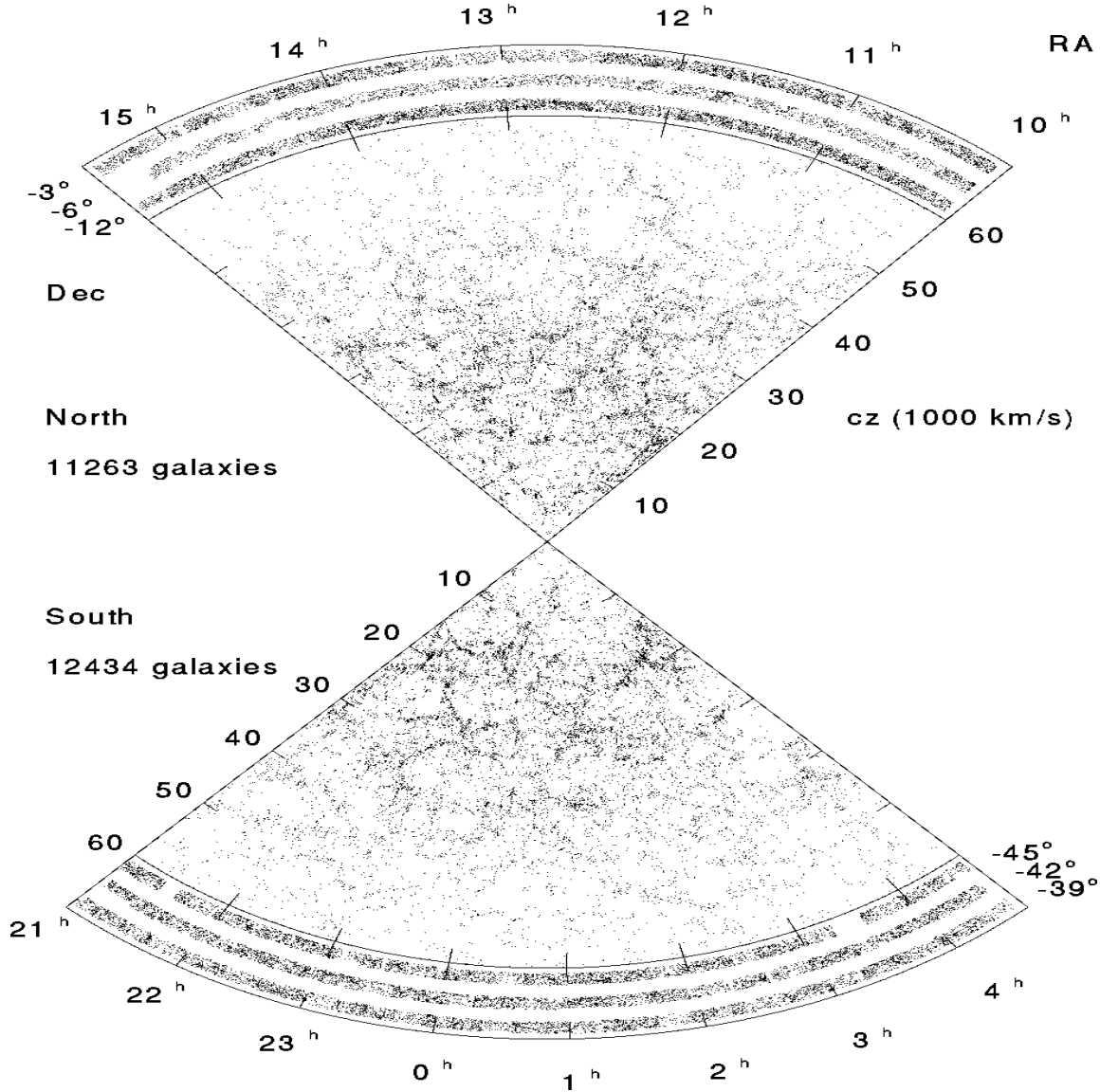, 6, 6
\caption{A redshift pie diagram of galaxies in the LCRS, kindly
supplied by Huan Lin.  The sky distribution of this survey is shown
below in Fig.~\ref{fig:surveys}.  The three slices in the Northern and
Southern Galactic caps are each shown projected on top of one
another. }
\label{fig:pie}
\end{figure}

  These surveys are best viewed not as contour plots as in
Fig.~\ref{fig:sgp}, but rather in the form of pie diagrams.
Fig.~\ref{fig:pie} shows the redshift distribution in the LCRS survey;
in each segment of the pie, three of the $1.5^\circ$ slices are
plotted on top of one another.  The angular coordinate is right
ascension, and the radial component is redshift.  Perhaps the most
striking feature of this map is the fact that one does {\it not\/} see
coherent structures stretching across the survey volume (compare, for
example, with the famous CfA2 slice of de Lapparent, Geller, \& Huchra
1986).  Bob Kirshner of the LCRS team has called this ``the end of
greatness'', in the sense that surveys are now probing a volume
appreciably larger than the largest structures in the universe.
We discuss below quantitative measures of structure on the largest
scales with this survey. 

  A similar survey has been carried out by Zucca \etal\ (1996) (cf.,
Guzzo 1996), using a fiber optic spectrograph on the ESO 3.6m
telescope; galaxies are selected from the Edinburgh-Durham Southern
Galaxy Catalogue (EDSGC; e.g., Heydon-Dumbleton, Collins, \&
MacGillivray 1989), and redshifts have been obtained for 3348 galaxies
to $b_J = 19.4$ over 36 square degrees.  This survey is called the ESO
Slice Project, or ESP.

  Outside the scope of this review are the very exciting redshift
surveys being done over small areas but going very deep, which attempt
to look for evolution in the properties of the galaxies themselves and
the large-scale structure they trace.  Lilly \etal\ (1995) have
carried out perhaps the most extensive redshift survey of faint
galaxies; their survey contains redshifts for 591 galaxies over five
small fields with $I < 22.5$; the median redshift is 0.56.  Still
deeper surveys have been done by Cowie, Broadhurst, Ellis and their
collaborators (cf., the review by Ellis 1996).

  Finally, there are two redshift surveys currently in the advanced
planning stages: the Sloan Digital Sky Survey (SDSS), which plans to obtain
redshifts for $10^6$ galaxies and $10^5$ quasar candidates over
$\approx 3.2$ ster in the Northern Galactic Cap to $r' \approx 18$,
and the Two-Degree Field Survey (2dF), which will go slightly deeper
to measure $250,000$ galaxies over a more limited region of sky.  We
will discuss these surveys in some detail below in \S~\ref{sec:future}. 

\section{The Luminosity and Selection Function}
\label{sec:phi(r)}
  As we emphasized in the previous section, in order to do
quantitative work with redshift surveys, one needs to know the
selection function of the sample.  As is apparent in
Fig.~\ref{fig:pie}, in a flux-limited sample, the number density of
galaxies drops off as a function of distance from the observer, which
is an effect we need to correct for.  Let us understand the issues for
the simplest case, that of a sample of galaxies limited to flux
$f_{min}$ completely and uniformly selected over a given region (cf.,
Sandage, Tammann, \& Yahil 1979; Efstathiou, Ellis, \& Peterson 1988;
Yahil \etal\ 1991; SW).  We define the {\it luminosity function}
$\Phi(L)$ of galaxies such that $\Phi(L)\,\d L$ is the average number
density of galaxies with luminosities (in some given band) between $L$
and $L + \d L$.  Let us make the assumption of a {\it universal\/}
luminosity function; that is, the number density of galaxies at
position \bfr\ with luminosity between $L$ and $L + \d L$ is a separable
function of \bfr\ and $L$: $\delta(\bfr) \Phi(L)\,\d L$.  Thus, we
assume that the luminosity function is independent of environment
(we'll discuss briefly in \S~\ref{sec:bias} observational evidence
that this does not hold true in great detail).

  With this assumption, it is straightforward to write down an
expression for the expected number density of galaxies in a redshift
survey at distance $r$, $n(r)$, in a universe without clustering: 
\begin{equation} n(r) = \int_{4\,\pi r^2 f_{min}}^\infty \Phi(L) \,
\d L\quad.
\label{eq:nr} 
\end{equation}

This suggests a simple estimator for the observed fractional
overdensity of galaxies $\delta(\bfr) \equiv (\rho(\bfr) -
\vev{\rho})/\vev{\rho}$ smoothed with a window $W(x)$:
\begin{equation} 
\delta(\bfr) = {1 \over \int \d^3\bfr\, W(r)} \sum_{\hbox{galaxies
$i$}} {W\left(|\bfr - \bfr_i|\right) \over n(r_i)} - 1\quad,
\label{eq:delta_r} 
\end{equation}
where $n(r)$ is given by Eq.~(\ref{eq:nr}) and $r_i \equiv |\bfr_i|$,
of course.  What this means, in effect, is that for the purposes of
deriving the density field $\delta$, each galaxy in the sample is
assigned a weight given by the inverse of $n(r)$ at the distance of
that galaxy.

  What is the best way to calculate $n(r)$?  It is straightforward
given a fit to the luminosity function.  There is a lengthy literature
of determinations of the luminosity function, which is reviewed in the
various articles listed at the beginning of this section. The classic
method, reviewed in, e.g., Felten (1977), simply involves binning the
galaxies in a redshift survey by absolute luminosity and dividing each
by the effective volume probed at that luminosity.  However, this
method gives unbiased results only in the limit that the number of
galaxies per unit redshift is unaffected by clustering, which is
exactly the quantity we ultimately wish to measure.  Here we present a
method of determining the luminosity function which does not require
any assumption about the large-scale homogeneity of galaxies; its
history can be traced through Sandage \etal\ (1979), Nicoll \& Segal
(1983), Efstathiou \etal\ (1988), Saunders \etal\ (1990), and Yahil
\etal\ (1991).  We will take a maximum likelihood approach and ask,
{\it given\/} that we know that a given galaxy $i$ has a distance
$r_i$, what is the likelihood that it have its observed flux between
$f_i$ and $f_i + \d f$?  The answer clearly depends on the luminosity
function, thus this approach will give us a handle on the luminosity
function itself. Indeed, this likelihood ${\cal L}_i$ is given by the
luminosity function at $L_i = 4\pi r_i^2 f_i$, normalized by the
integral of the luminosity function over all luminosities it could
have, given the flux limit. Formally:
\begin{equation} 
{\cal L}_i = {{ \Phi(L_i) \,\d L} \over { \int_{4\pi r_i^2
f_{min}}^\infty\Phi(L_i) \,\d L}} \propto {{\d n(r)/\d r \Big|_{r = r_i
(f_i/f_{min})^{1/2}}}\over {n(r_i)}}\quad,
\label{eq:likelihood} 
\end{equation}
where the proportionality follows directly from Eq.~(\ref{eq:nr}).
The constant of proportionality is just $\d L/(8\,\pi r f_i)$, which
is independent of the parameters of the 
selection function, and therefore does not concern us as we maximize the
likelihood below.  
Given the likelihood for observing any one galaxy, the likelihood for
an entire sample is given by the product of this expression for all
galaxies in the sample.  The maximum likelihood method then consists
of the following: choose a parametric form for $n(r)$.  Calculate the
likelihood function over all the galaxies in the sample as a function
of the parameters, and find its maximum (as is often done in this sort
of exercise, the quantity actually maximized in practice is the
logarithm of the likelihood function).  This determines $n(r)$, from
which the luminosity function follows as a simple derivative.  A
variant of this approach involves specifying the luminosity function
not in terms of a smooth functional form, but as a series of constants
in bins. Nicoll \& Segal (1983), Efstathiou \etal\ (1988), and Koranyi
\& Strauss (1996) show how the likelihood can be maximized with
respect to the values of the constants through a straightforward
iterative procedure.

  It is clear how the likelihood procedure outlined above is
independent of density inhomogeneities: the {\it positions} of
galaxies are taken as a prior, and one asks for the likelihood of
observing their fluxes.  Thus the results are not biased by the
distribution of the positions of the galaxies.  But a consequence of
this is that the {\it normalization} of $\Phi(L)$ (or equivalently, of
$n(r)$) is not determined.  With this in mind, we explicitly drop the
normalization of $n(r)$ to define the {\it selection function}:
\begin{equation} 
\phi(r) \equiv \left\{ 
\begin{array}{ll}
{{ n(r)} \over {n(r_s)}}, &r > r_s \\
1&r \le r_s\quad, \\
\end{array}
\right.
\label{eq:selfunct} 
\end{equation}
where $r_s$ is some small fiducial distance, typically 500 \kms, below
which the selection function is set to unity.  That is, the selection
function quantifies the fraction of the luminosity function seen at a
distance $r$, relative to the numbers at $r_s$.  This of course begs
the question of how to normalize the selection function itself in
order to calculate $\delta(\bfr)$. The usual assumption is that the
entire volume covered by a given redshift survey is a fair sample of
the universe; that is, the mean density of galaxies in this volume
differs negligibly from the true mean density.  Given this, there are
several ways one can calculate the mean density as a weighted sum over
the galaxies in the sample (Davis \& Huchra 1982).  The simplest is to
define 
\begin{equation} 
n(r_s) = {1 \over V} \sum_{\hbox{galaxies $i$}} {1 \over \phi(r_i)}\quad,
\label{eq:n1} 
\end{equation}
where the sum is over all galaxies in the sample between $r_s$ and
some outer radius beyond which the sample gets too sparse to be
useful, and $V$ is the volume enclosed.  Davis \& Huchra (1982) derive
a minimum variance version of Eq.~(\ref{eq:n1}) that takes the known
clustering of galaxies into account, although in practice, it requires
knowing the strength of the correlation function on the largest
scales, where it is most poorly understood. Another approach is that
adopted by Saunders \etal\ (1990), who note that in the approximation
that the sample is uniform, the total number of galaxies in the sample
${\cal N}$ should be given by $\int\! \d^3 \bfr\, n(r)$ (where the integral
is over the solid angle and depth of the survey), which allows one
to normalize $n(r)$. 

  It is actually quite straightforward to modify the maximum
likelihood method sketched out here to the more general case of
selection depending on more complicated criteria, such as both a
magnitude and diameter cut, or extinction as a function of direction;
some of these complications are described in Santiago \etal\ (1996).
Another interesting issue discussed in that paper, and also in
Mancinelli (1996), is the effect of flux errors on the derived
luminosity function, selection function, and density field. The
derived luminosity function is given by the true luminosity function
convolved with the flux errors, not surprisingly.  This gives a
selection function which falls less rapidly with distance than in the
case without flux errors.  One might think that this could cause the
derived density field to be systematically biased as a function of
distance, but there is a competing effect which goes in the opposite
direction, namely Malmquist bias.  Flux errors scatter galaxies over the flux
limit of the sample.  Because the number of galaxies is a
monotonically decreasing function of flux, more galaxies scatter into
the sample than out of it.  If the flux errors are proportional to the
flux itself (such as one gets with magnitude errors which are constant
with magnitude), these two
effects cancel exactly; the derived density field is
unbiased, even though the luminosity function is biased\footnote{The
caveat that the flux errors must be proportional to flux is not made
clear in Santiago \etal\ (1996).  However, there is no condition on
the Gaussianity of the error distribution.}.

\section{Clustering Statistics}
\label{sec:clustering}

As Figs.~\ref{fig:sgp} and \ref{fig:pie} make clear, galaxies are not
uniformly distributed in space; indeed, one of the main motivations
for doing redshift surveys is to quantify the observed clustering and
from it, draw inferences about larger cosmological questions such as
are addressed throughout this volume: what is the nature of dark
matter? What is the value of the Cosmological Density Parameter? How
did galaxies form? We will be able to address only some of these
issues here.  But let us ask how best to quantify in a statistical way
the clustering that is seen here. 

 The current dominant paradigm for the formation of large-scale
structure postulates that it grows by the process of gravitational
instability from an early epoch when the fluctuations were of very low
amplitude (recent textbooks treating this include Kolb \& Turner 1990;
Peebles 1993; Padmanabhan 1993; Coles \& Lucchin 1995).  This basic
picture was given great support by the fact that the fluctuations in
the Cosmic Microwave Background (CMB) as detected by the Cosmic
Background Explorer (COBE; Smoot \etal\ 1991) are within a factor of
two of those expected from extrapolation of the present-day observed
clustering of galaxies (e.g., Wright \etal\ 1992; cf., the
contribution from Silk in this volume). Moreover, if the fluctuations
arise from inflationary processes in the very early universe (e.g.,
Kolb \& Turner 1990), they turn out to be {\it random phase}.  This
means the following.  Given the density field $\delta(\bfr)$ at some
early time, one can take its Fourier Transform to obtain the complex
quantity $\tilde \delta(\bfk) = \left|\tilde \delta(\bfk)\right| e^{i
\theta}$, where $\theta$ is the phase of each mode.  Inflationary
models predict the quantity $\theta$ to be uniformly distributed
between 0 and $2\,\pi$\footnote{Examples of models in which the phases
are not random include those with gravitating seeds (such as cosmic
string loops or textures) or explosion models. A general discussion of
the large-scale structure implications of these models can be found in
Weinberg \& Cole (1992).}.

  In any case, if the phases are random, then the Central Limit
Theorem implies that the distribution function of the density field is
Gaussian\footnote{However, the converse is not true: a Gaussian
distribution function for the density field does not imply random
phases.}.  What this means is that all its reduced moments of third
order and higher vanish (after all, they are defined relative to a
Gaussian).  That is, one can give a full statistical description of
the density field by specifying its second moment with respect to the
only relevant independent variable, the scale (the Cosmological
Principle says that space is isotropic, which is confirmed to
fantastic accuracy with the COBE data, so there is no directional
dependence to the clustering in a sufficiently large sample).  There
are a number of ways we might quantify this.  One way is simply to
calculate the statistic implied by this discussion: the second moment
of the density distribution as a function of scale.  That is, one asks
for the variance in $\delta$, averaged over spheres of a given radius
$r$; this quantity will be referred to as $\sigma^2(r)$.
Alternatively, one can calculate the {\it correlation function} of the
density field: $\xi(r) \equiv \vev{\delta(\bfx)\delta(\bfx + \bfr)}$,
where the averaging is over position $\bfx$ and over direction of
\bfr.  Finally, one can define the {\it power spectrum\/} of the
density field, which is the modulus squared of $\tilde\delta$:
\begin{equation} 
\vev{\tilde \delta(\bfk)\tilde \delta^*(\bfk')} = (2\,\pi)^3 P(k) \delta_D(\bfk -
\bfk')\quad,
\label{eq:pk-def} 
\end{equation}
where the averaging is over directions of \bfk, and $\delta_D$ is the
Dirac $\delta$ function\footnote{The normalization, and even the
definition of $P(k)$ depends on one's Fourier Transform convention in
defining $\tilde \delta$; compare Eq.~(\ref{eq:pk-def}) with that in
Peebles (1980), for example.}.

These three statistics
are related to each other in straightforward ways:
\begin{equation} 
\xi(r) = {1 \over 2\,\pi^2} \int\! \d k\, k^2 P(k) {\sin kr \over kr}
\label{eq:xi-Pk} 
\end{equation}
and
\begin{equation} 
\sigma^2(r) = {1 \over {2\,\pi^2}} \int\! \d k\, k^2 P(k) \widetilde W^2(kr)\quad,
\label{eq:sigma-Pk}
\end{equation}
where $\widetilde W(x)$ is given by a spherical Bessel function for
a spherical window:
\begin{equation} 
\widetilde W(x) = {{3(\sin x - x \cos x)} \over x^3}\quad.
\label{eq:window} 
\end{equation}

  How does this relate to the galaxy distribution?   As we said
before, we believe (although do not have definitive proof) that
clustering grew through the process of gravitational instability.
In linear perturbation theory it is straightforwardly shown (cf., the
textbooks referenced above) that the amplitude of perturbations grows
as a function of time, independent of the wavelength of the
perturbations. What this means is that the power spectrum and related
statistics change in {\it amplitude}, but not in {\it shape}, as
perturbations grow. Therefore, measurements of the shape of the power
spectrum as a function of $k$ today are a direct measure of its shape
in the past.  That prospect is very exciting, because the shape of the
power spectrum is commonly modelled to be due to two components. One
is the {\it primordial\/} power spectrum, i.e., that laid down,
perhaps during inflation, when the universe was very young. This is
modified by the differing growth of perturbations of super-horizon and
sub-horizon scales before and after the epoch of matter-radiation
equality (cf., Kolb \& Turner 1990; Efstathiou 1991 for reviews). The
details of this are determined by the nature of the dark matter (i.e.,
hot or cold) and how much of it there is (as this determines the
matter-radiation equality epoch).  

Thus in principle, measurements of the galaxy power spectrum can tell
us a great deal about the early universe and the dark matter.
However, there are some real complications that come in.  The first is
that we are observing galaxies, while it is the distribution of dark
matter that the theories predict.  Thus we need a model for the
relative distribution of galaxies and dark matter.  The simplest
assumption is that the distribution of galaxies and dark matter are
the same (i.e., $\delta_{\rm galaxies}(\bfr) = \delta_{\rm dark\
matter}(\bfr)$), but it was realized in the mid-1980's (Kaiser 1984;
Davis \etal\ 1985; Bardeen \etal\ 1986; Dekel \& Rees 1987) that there
is no {\it a priori\/} reason that this might be true.  Indeed, one
could explain a number of observations (such as the relative strength
of the cluster and galaxy correlation functions, and the pairwise
velocity dispersion of galaxies on small scales) if it were false. The
simplest model of biasing has that the galaxy and dark matter density
fields differ by a constant factor $b$. That is,
\begin{equation} 
\delta_{\rm galaxies}(\bfr) = b\,\delta_{\rm dark\ matter}(\bfr)\quad,
\label{eq:linear-bias} 
\end{equation}
independent of smoothing scale, although reality is almost certainly
more complicated than this. In any case, to the extent that this {\it
linear biasing} holds, the shape of the derived galaxy power spectrum
of galaxies will still be the same of that of the dark matter.

  The next complication is due to non-linear evolution of the power
spectrum. The statement that density perturbations grow at a rate which
is independent of scale holds only when the perturbations are in the
linear regime, i.e., $\delta \ll 1$.  When this condition no longer
holds, all modes no longer grow independently, and the growth rate
does indeed become a function of scale, meaning that the non-linear
power spectrum no longer keeps the same shape as its linear
progenitor. In practice, because density perturbations are generically
an increasing function of $k$ (at least for power spectra $P(k)
\propto k^n, n \ge -3$, cf., Eq.~\ref{eq:sigma-Pk}), this means that
the power spectrum is modified by non-linear effects on small
scales. To a certain degree the growth of the power spectrum on small
scales can be calculated analytically (e.g., Jain \& Bertschinger
1994) or phenomenologically (Hamilton \etal\ 1991; Jain, Mo, \&
White 1995), and indeed there is a quite extensive literature on
extensions of linear theory for the growth of perturbations into the
non-linear regime (cf., the review by Sahni \& Coles 1995). 

As we will see below, the power spectrum measured within the
necessarily finite volume of any given redshift survey is not
identical to the theoretical ideal of that measured in an infinite
volume, essentially because of the difficulties of defining the
continuous Fourier Transform in a finite volume.  Indeed, the measured
power spectrum is a convolution of the ``true'' power spectrum with
the Fourier Transform of the observing volume, which tends to depress
the power spectrum on large scales. 

  Finally, what we observe for each galaxy in a redshift survey is a
redshift, not a distance. Only in the approximation that peculiar
velocities are negligible are the two the same (cf.,
Eq.~\ref{eq:cz-correct}).  Peculiar velocities have two effects on the
power spectrum as measured in redshift space. On small scales, the
pairwise velocity dispersion of galaxies spreads galaxies out in
redshift space relative to their distribution in real space (think of
a cluster of galaxies stretched out into a ``Finger of God'' in
redshift space), thereby decreasing the apparent amplitude of
clustering on small scales.  On large scales, the dominant effect is
due to coherent streaming of galaxies towards overdensities, giving a
compression in redshift space, and therefore amplifying the apparent
clustering (Kaiser 1987).

With all these effects acting, the interpretation of the observed power
spectrum is thus quite non-trivial, and the next section of this
review concentrates on the details of how people have tried to take
these various effects into account, and indeed, to take advantage of
them to get additional information out of the available data. 

\section{Measurements of the Power Spectrum}
\label{sec:power-spectrum}

\subsection{Techniques}
\label{sec:Pk-techniques}

For two decades, the standard way to quantify the clustering seen in
galaxy surveys was through the use of the $N$-point correlation
functions, especially the two-point correlation function $\xi(r)$
(Peebles 1980; SW).  However, much of the recent developments in the
field have been on analyses of its Fourier Transform $P(k)$ (cf.,
Eq.~\ref{eq:xi-Pk}), and we will emphasize this in this review.

The Fourier Transform of the galaxy density field (Eq.~\ref{eq:delta_r}) is
\begin{equation} 
\tilde\delta({\bf k}) = {{1}\over{nV}}
\sum_i {{1}\over{\phi(r_i)}} e^{\mathrm{i}\bfk\cdot\bfr_i} -
W(\bfk)\quad ,
\label{eq:delta-discrete-k}
\end{equation}
where
\begin{equation} 
W(\bfk)\equiv {{1}\over{V}}\int_V \d^3\bfr\,
e^{\mathrm{i}\bfk\cdot\bfr}
\label{eq:window-ft}
\end{equation}
is the Fourier Transform of the survey volume. 
Our estimator of the power spectrum is then
\begin{equation} 
\Pi ({\bf k})\equiv V\tilde\delta(\bfk)\, \tilde\delta(\bfk)^\ast\quad,
\label{eq:power-est}
\end{equation}
where the factor of $V$ on the right hand side gets the units right.
Several lines of algebra (e.g., Fisher \etal\ 1993; Vogeley 1995) show that the
expectation value of this estimator is given by
\begin{equation}
\vev{ \Pi (\bfk)} =  \int \d^3\bfk' \,P(k')G(\bfk-\bfk')
+{{1}\over{nV}}\int \d^3\bfr{1\over{\phi(r)}}\quad,
\label{eq:Pi-power}
\end{equation}
where
\begin{equation} 
G(\bfk-\bfk')\equiv {{V}\over{(2\pi)^3}}|W(\bfk-\bfk')|^2\quad.
\label{eq:convolving-window}
\end{equation}
In the limit of an infinitely large volume, $G$ approaches a Dirac
delta function, as it must.  Thus the power spectrum estimator is
given by the true power spectrum convolved with an expression
involving the Fourier Transform of the volume, plus a shot noise
term. Because one normalizes the density field assuming the mean
density inside the surveyed volume is equal to the true global mean,
there is an additional correction factor to Eq.~(\ref{eq:Pi-power}) to
compensate for the resulting loss of power; this term is important for
measurements of the power spectrum on scales approaching that of the
survey itself.

The quantity in Eq.~(\ref{eq:Pi-power}) is still a function of \bfk,
and thus must be averaged over solid angle in \bfk-space in order to
calculate a quantity dependent only on $k$.  This is a straightforward
procedure for values of $k$ probing scales appreciably smaller than
the survey dimensions.  However, when $1/k$ becomes comparable to the
smallest dimension of the survey, this averaging can mix together
modes with very different convolutions with the survey window (simply
because $G$ in Eq.~(\ref{eq:convolving-window}) becomes quite
anisotropic for surveys with restricted geometries).  A related
problem is that an appreciable covariance can develop between
determinations of the power spectrum for different values of
$k$. Different workers have found different ways of dealing with these
problems.  Fisher \etal\ (1993), working with the \iras\ 1.2 Jy
survey, have close to full sky coverage, and therefore are not
affected much by the anisotropy of the window function.  They measure
$\Pi(\bfk)$ within cylinders embedded within the survey volume, whose
long axis of length $2\,R$ is parallel to the vector \bfk. Choosing
$kR$ to be an integral multiple of $\pi$ means that
$\delta(\bfk)$ now scales exactly with the mean density, and thus
errors in the mean density affect only the {\it amplitude}, and not
the shape, of $P(k)$. 

Feldman, Kaiser, \& Peacock (1994) approach the problem from a
different viewpoint, by asking for a weighting to
Eq.~(\ref{eq:delta-discrete-k}) that allows them to measure $P(k)$
with the minimum variance. For the case of a full-sky sample, and
assuming that the different Fourier modes have random phases, they
derive the weights: 
\begin{equation} 
w_i = {1 \over {1 + n \phi(r_i) P(k)}}\quad.
\label{eq:power-weight} 
\end{equation}
 With this weight function, the variance in
the estimate of the power spectrum is given by
\begin{equation} 
\sigma^2[P(k)] = {{(2 \,\pi)^3} \over  {V_k  \int\! \d^3 \bfr
\left[n w \phi(r)\right]^2}}\quad,
\label{eq:power-error} 
\end{equation}
where $V_k$ is the volume in $k$-space occupied by the bin in
question. This expression assumes that the bins are spaced far enough
apart that the covariance is negligible (this happens roughly for
separations $\Delta k > 2\,\pi/R$, where $R$ is the characteristic
dimension of the volume surveyed).

  Tegmark (1995) has carried this type of analysis further, by asking
for the optimal weighting of $\Pi(\bfk)$ as a function of
amplitude and direction of \bfk, given the survey geometry.  One wants
to measure the power spectrum with as much resolution in $k$ as
possible, without introducing large amounts of covariance between
adjacent values.  Tegmark shows that this is done with a weighting
function $w(\bfr)$, which
is the ground-state solution to the Schr\"odinger equation with
potential given by the inverse of the selection function:
\begin{equation} 
\left[-{1\over 2}\nabla^2 + {\gamma \over\phi(\bfr)}\right]w(\bfr) =
E\,w(\bfr)\quad,
\label{eq:schrodinger} 
\end{equation}
where $\gamma$ is a parameter which determines the resolution in $k$
of the determination of the power spectrum (at the expense of
signal-to-noise ratio). 

  Vogeley \& Szalay (1996) and Tegmark, Taylor, \& Heavens (1996) have
addressed this problem at one further level of sophistication.  While
the Fourier modes are the ideal basis in which to expand the density
field in the ideal case of an infinite survey, this is not the case in
the realistic case of a survey covering a finite area of sky, with a
radial selection function.  In particular, the Fourier modes are not
orthonormal over the survey volume.  Thus the above authors expand the
observed density field in orthonormal eigenmodes which maximize the
signal-to-noise ratio, given the survey geometry and selection
function, using the Karhunen-Lo\`eve (K-L) transform.  These modes are
linear combinations of the standard Fourier modes which enter the
power spectrum.

  Following Vogeley \& Szalay, divide a survey volume into a series of
$M$ volume elements centered at positions $\bfx_i$ with volumes
$V_i$.  Let $f(\bfx_i)$ be the counts of galaxies observed within cell
$i$. 
We expand the $f(\bfx_i)$ in a series of orthonormal basis vectors
$\pmb{$\Psi$}_j$, i.e.,
\begin{equation} 
f(\bfx_i) = \sum_j \pmb{$\Psi$}_j(\bfx_i)\, B_j\quad .
\label{eq:K-L-expansion} 
\end{equation}
The K-L transform uses
the basis vectors which satisfy the eigenvalue problem:
\begin{equation} 
{\bf R} \pmb{$\Psi$}_j = \lambda_j \pmb{$\Psi$}_j\quad,
\label{eq:eigen-equation} 
\end{equation}
where {\bf R} is the correlation matrix of the $f$'s: $R_{ij} =
\vev{f(\bfx_i)f(\bfx_j)}$, and the eigenvalues are $\lambda_j =
\vev{B_j^2}$. The relation to the power spectrum is clear: the matrix
$R_{ij}$ has elements given by the sum of the correlation function
between volume $V_i$ and $V_j$, and contributions from shot
noise. Given the K-L expansion of an observed density field, the
best-fit power spectrum can be found by the methods of maximum
likelihood. 

Vogeley \& Szalay point out that the K-L transform is a maximally
efficient representation of the data, in the sense that if the
eigenmodes are ordered in decreasing order of their eigenvalues, the
truncation of the eigenvectors to the first $N$ gives a
representation of the data that differs from the truth by as small
an amount as possible. Another way to say this is that this gives an
expansion of the data in modes of decreasing signal-to-noise ratio.  A
dramatic demonstration of the power of the K-L transform to compress
data, in quite a different astronomical context, is that of Connolly
\etal\ (1995), who show that the optical spectral energy
distributions of galaxies can be well represented by their first three
eigenvectors. 

This approach to analysis of redshift surveys promises the greatest
advantage over the standard power spectrum analysis in the case of
surveys with sharp boundaries, especially those covering a fraction of
the sky with anisotropic sky coverage, such as slice surveys or pencil
beam surveys.  The LCRS is an example of a slice survey. It has not
yet been analyzed using the methods of the K-L Transform (although I
understand that Vogeley \& Szalay intend to do so), but Landy \etal\
(1996) do an analysis of the power spectrum of this survey in the same
spirit.  Their survey consists of six narrow slices, and thus
recognizing that there is little information on the large-scale power
spectrum in a direction perpendicular to the slice, they calculate in
effect the two-dimensional analogue of the power spectrum on the
density field of the slices, collapsed along this narrowest
direction. 

\subsection{Results}
\label{sec:Pk-results}

The power spectrum of various redshift surveys of galaxies has been
measured by a large number of groups (Baumgart \& Fry 1991; Peacock \&
Nicholson 1991; Park, Gott, \& da Costa 1992, Vogeley \etal\ 1992;
Fisher \etal\ 1993, Feldman \etal\ 1994; Park \etal\ 1994; da Costa
\etal\ 1994b; Tadros \& Efstathiou 1995, 1996; Vogeley 1995; Landy
\etal\ 1996; Lin \etal\ 1996).  A summary of results through 1994 is
given by Peacock \& Dodds (1994), and is reviewed in SW.  For the most
part, this large range of determinations of $P(k)$ is remarkable for
the uniformity of the results they give. The amplitude of the power
spectrum determined from different surveys often differs
significantly, even for galaxies selected in the same way (compare
Fisher \etal\ 1993, Feldman \etal\ 1994, and Tadros \& Efstathiou 1995
for \iras-selected galaxies), but there seems to be broad agreement
between groups on the shape of the power spectrum, at least within the
rather large error bars.  This is especially remarkable, given the
very different geometries, and therefore window functions, of the
different surveys (cf., Eq.~\ref{eq:Pi-power}). The observed power
spectrum is a power law, $P(k) \propto k^n, \ n \approx -2$ on small
scales, changing to $n \approx -1$ on scales above $\lambda \equiv
2\,\pi/k \sim 30\mpc$.  There is tentative evidence for a flattening
of $P(k)$ on the largest scales on which it has been measured,
$\lambda \sim 100-200\mpc$.  The best evidence for this comes not from
a redshift survey, but rather the APM photometric survey of $2 \times
10^6$ galaxies over the Southern Galactic Cap (Maddox \etal\ 1990abc,
1996). Baugh \& Efstathiou (1993, 1994; cf., Gazta\~naga 1995) discuss
the determination of the spatial power spectrum from the angular
correlation function; the evidence for a flattening of the power
spectrum is fairly unambiguous. 

\begin{figure}
\fig 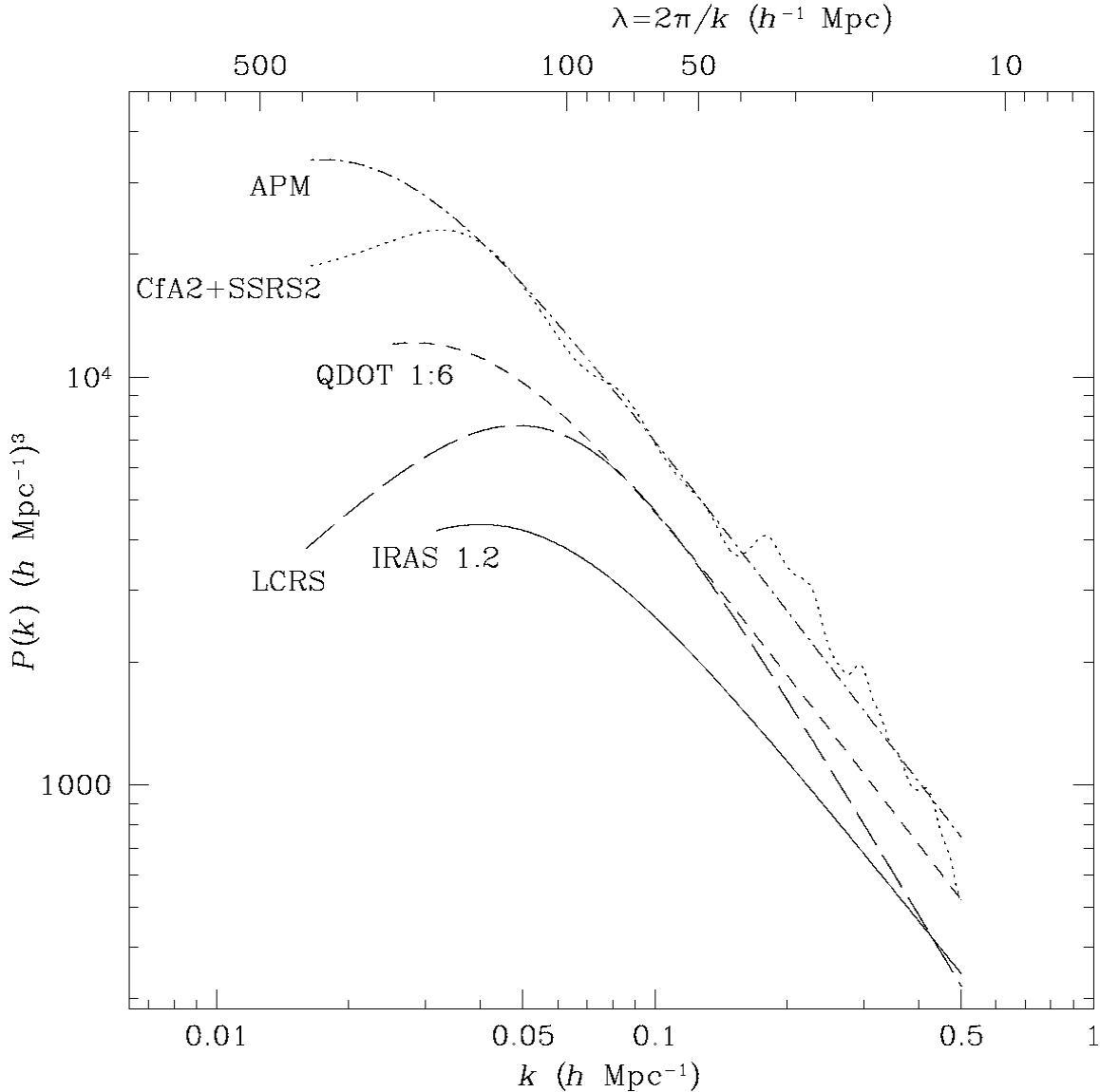, 6, 6
\caption{A real space correlation functions of some of the surveys
discussed in the text, after a similar figure in Kolatt \& Dekel
(1995).  The CfA2+SSRS2 line is drawn from interpolation between
datapoints given in da Costa \etal\ 1994b; the others are
parameterized fits to the observed power spectra. The APM curve is
from Baugh \& Efstathiou (1993), QDOT from Feldman \etal\ (1994), LCRS
from Lin \etal\ (1996), and \iras\ 1.2 from Fisher\etal\ (1993).  Each
curve extends to the largest scales on which the power spectrum was
determined.}
\label{fig:kolatt-power}
\end{figure}

  The power spectra derived from the surveys described above are shown
in Fig.~\ref{fig:kolatt-power}, which corrects the published power
spectra in redshift space for distortions as predicted in linear
theory using Eq.~(\ref{eq:kaiser1}) below, following Kolatt \& Dekel
(1995).  With the exception of the CfA2+SSRS2 curve, all curves shown
are fits of the power spectra to simple functional forms.  Thus this
figure does not give a sense of the size of the error bars (compare
with Fig.~\ref{fig:sdss-power} below).  The amplitudes of the
different power spectra are indeed quite different from one
another\footnote{Tadros \& Efstathiou (1995) go a long ways towards
reconciling the amplitudes of the two \iras\ surveys in
Fig.~\ref{fig:kolatt-power}.}, varying by almost a factor of 3 at $k =
0.1 h\,\rm Mpc^{-1}$, but as remarked above, the shapes are remarkably
the same.

  Before 1992, theoretical predictions for the form of the power
spectrum had an important freedom.  Current models for the origin of
density fluctuations invoking quantum processes in the early universe do not
constrain the amplitude of these fluctuations (essentially because we
do not yet have a detailed enough model of the relevant particle
physics).  Thus the normalizations of the power spectra were
essentially unconstrained. However, now with observations of the
fluctuations of the CMB with COBE (cf., Bennett \etal\ 1996;
G\'orski \etal\ 1996 for the latest results), this normalization is
tied down to 10\% accuracy for any given model\footnote{This is
true only to the extent that the fluctuations are interpreted as being
due solely to the Sachs-Wolfe (1967) effect on large scales.  However,
there exist models in which a substantial contribution to the
fluctuations comes in the form of gravitational waves (e.g., Davis
\etal\ 1992), making the relationship between the COBE fluctuations
and the normalization of the power spectrum
ambiguous.} (cf., Bunn \& White 1996 and references therein).

In any case, we can now make definite predictions, including
normalization, for the power spectra of different models.  If the
biasing of galaxies relative to mass is independent of scale on large
scales, then the measured galaxy power spectrum definitely rules out
the standard Cold Dark Matter model, as defined, e.g., in Davis \etal\
(1985): adiabatic fluctuations with primordial spectral index $n = 1$,
$\Omega = 1$, and $H_0 = 50 \kms\rm\,Mpc^{-1}$. If biasing is {\it
local}, i.e., if the probability that a galaxy form at a given place
is a function of the physical properties of that region of space
within a few Mpc, then the bias is indeed probably independent of
scale (Weinberg 1995; Kauffmann, Nusser, \& Steinmetz 1996); if this
assumption does not hold, one can find models in which standard CDM
fits the observed power spectrum on large scales (Bower \etal\ 1993).
Leaving this last possibility aside, the observed power spectrum with
COBE normalization can be fit by a range of variants on the standard
CDM model, all designed to give more power on large scales than small:
moving the epoch of matter-radiation equality later by changing the
value of $\Gamma = \Omega h$, by decreasing the index of the
primordial power spectrum by a few tenths, and/or by replacing some of
the CDM with Hot Dark Matter (cf., the contributions from Primack and
Ostriker in this volume).

As mentioned above, there is rough agreement between different workers
as to the shape of the galaxy power spectrum on intermediate and large
scales.  There is one dramatic exception to this.  Landy \etal\
(1996), in their analysis of the two-dimensional power spectrum
from the LCRS, find evidence for a strong peak in the power spectrum
on the largest scales probed ($2\,\pi/k \approx 100\mpc$), rising a
factor of several above the best-fit CDM-like power spectrum as
measured on smaller scales.  This result, if true, is very important,
and may point to baryon-dominated isocurvature models (Peebles 1987)
which are designed to have a bump in the power spectrum on these
scales.  However, it remains unclear how such a bump could exist, and
yet not appear in the angular power spectrum of Baugh \& Efstathiou (1993,
1994).

\subsection{Redshift-Space Distortions}
\label{sec:redshift-distortions}

\subsubsection{Linear Scales}
\label{sec:linear-redshift}

We remarked above that measurements of the power spectrum in redshift
space differ systematically from those in real space, due to the
distorting effects of peculiar velocities.  On large scales
(i.e., those on which density perturbations are small), a linear
perturbation expansion of the equations of gravitational instability 
yields a simple relation between the peculiar velocity and density
fields (Peebles 1980):
\begin{equation} 
\nabla \cdot \bfv(\bfr) =  -\Omega^{0.6} \delta(\bfr)
\label{eq:linear-theory} 
\end{equation}
(cf., the contribution by Dekel in this volume).  Given
this, one can calculate that the 
effect of coherent infall into overdensities is to multiply the
Fourier modes by a fixed operator independent of scale (Kaiser 1987):
\begin{equation} 
\tilde \delta(\bfk) \Rightarrow (1 + \beta \mu^2)\tilde \delta(\bfk)\quad; 
\label{eq:kaiser1} 
\end{equation}
in real space, this becomes the operator equation:
\begin{equation} 
\delta(\bfr) \Rightarrow \left[1 + \beta \left({\partial \over
\partial z}\right)\nabla^{-2}\right] \delta(\bfr)\quad, 
\label{eq:kaiser1_real} 
\end{equation}
where $\beta \equiv \Omega^{0.6}/b$ is the proportionality constant
between the {\it galaxy\/} density field and the divergence of the peculiar
velocity field (cf., Eq.~\ref{eq:linear-bias}), and
$\mu$ is the cosine of the angle between the wavevector \bfk\ and the
line of sight. It is a straightforward calculation to propagate the
effects of Eq.~(\ref{eq:kaiser1}) through to the power spectrum; the
power spectrum gets multiplied by a factor:
\begin{equation} 
K(\beta) = \left (1 + {2 \over 3} \beta + {1 \over 5} \beta^2\right)\quad,
\label{eq:kaiser2} 
\end{equation}
which is an appreciable correction; $K(\beta = 1) = 1.87$. 

On small scales, linear theory breaks down, and peculiar velocities
are dominated by the pairwise velocities of galaxies in groups and
clusters. The effect of this, as we saw above, is to reduce the
amplitude of clustering in redshift space relative to real space.  As
we will see in a moment, the amplitude of the small-scale velocity
dispersion of galaxy pairs remains quite unclear, making it difficult
to predict {\it a priori\/} on what scales the redshift distortions
make the transition from nonlinear to linear behavior (cf., the
discussion in Brainerd \& Villumsen 1993; Gramann, Cen, \& Bahcall
1993; Fisher \etal\ 1994b).  These effects can measured by the fact
that they make the clustering in redshift space anisotropic; that is,
the radial direction is the only one affected by distortions.  This
can be seen directly by measuring the correlation function, not as a
function simply of redshift space separation, as is usually done, but
rather as a function of the separation both perpendicular and along
the line of sight; redshift space distortions will make the contours
of $\xi$ in this space anisotropic. This has been carried out by a
number of workers with various datasets (Hamilton 1993, Fisher \etal\
1994b; Loveday \etal\ 1996).  One can carry out similar analyses on
the power spectrum (Cole, Fisher, \& Weinberg 1994, 1995; Fisher \&
Nusser 1995; Taylor \& Hamilton 1996; Hamilton 1996), or on a
spherical expansion of the density field (Fisher, Scharf, \& Lahav
1994c; Heavens \& Taylor 1995); these various approaches are reviewed
in SW.  All these results have been limited by the fact that existing
datasets still do not cover enough volume that the underlying
real-space clustering on large scales can be adequately modelled as
isotropic; the surveys cover only a relatively small number of
superclusters, whose orientations relative to the line of sight do not
necessarily average out. This is reflected in the range of values of
$\beta$ and the errors that people quote in their detection of this
effect; most workers detect the distortions due to peculiar velocities
on large scales at only the 3-$\sigma$ level or so.  Results from the
correlation function and power spectrum redshift anisotropies give
$\beta = 0.4 - 0.7$, with large error bars 1/3 as large as the signal;
interestingly, the spherical expansion method quoted above gives
appreciably larger values, $\beta = 1$ for the same data.  In any
case, the surveys discussed in \S~\ref{sec:future} promise to survey a
large enough volume to give much smaller error bars, and should be
among the most exciting science done with these samples.

There is another technical problem in these analyses. On the face of
it, Eq.~(\ref{eq:kaiser1}) seems nonsensical; what does $\mu$, the
cosine of the angle between the line of sight (defined in real space,
of course) and the vector \bfk\ mean?  One is using the ``distant
observer'' approximation, wherein the sample is approximated to be far
from the observer, allowing the quantity $\mu$ to be defined (cf., the
discussion in Cole \etal\ 1994). This means that in practice, one must
either work with a fraction of the galaxy pairs available (especially
in a full-sky redshift survey, where the distant observer
approximation is most grossly violated), or work with a quantity which
is related to the theorist's ideal in a very complicated way. Zaroubi
\& Hoffman (1996) show that even in the limit of an infinite universe,
the Fourier modes of the density field measured in redshift space are
coupled. Recently, Hamilton \& Culhane (1996) have suggested a
generalization of Eq.~(\ref{eq:kaiser1_real}) that allows not having
to invoke the distant observer approximation:
\begin{equation} 
\delta(\bfr) \Rightarrow \delta(\bfr)\left[1 + \beta\left({\partial^2
\over \partial r^2} + {\partial \ln r^2 \phi(r) \over \partial \ln r}
{\partial \over r \partial r}\right)\nabla^{-2}\right]\delta(\bfr)\quad,
\label{eq:hamilton} 
\end{equation}
which is an operator equation both in real space (as shown here) and
$k$-space (Zaroubi \& Hoffman 1996). This approach has yet to be
applied to real data.

\subsubsection{Non-linear Scales}
\label{sec:nonlinear-redshift} 

On small scales, it is the pairwise velocity dispersion of galaxies,
especially in virialized systems, that dominates the redshift space
distortions.  The small-scale velocity dispersion of galaxies is
interesting for two reasons: first, given various assumptions about
the stability of galaxy pairs on small scales, it can be related to
the two-point and three-point correlation functions and to the value
of $\Omega$ (the Cosmic Virial Theorem) (Peebles 1976ab, 1980; cf.,
the discussion in Fisher \etal\ 1994b; Bartlett \& Blanchard 1995;
Kepner, Summers \& Strauss 1996).  Second, it is a quantity that can
be predicted for various cosmological models from $N$-body
simulations, and has been used in the literature in the past as a
strong, although not completely unambiguous, discriminator between
models (for the standard CDM model alone, which seems to overpredict
the small-scale velocity dispersion by a large factor, one can follow
the controversy through Davis \etal\ 1985; Couchman
\& Carlberg 1992; Cen \& Ostriker 1993; Brainerd \& Villumsen 1994; Zurek \etal\ 1994; 
and Brainerd \etal\ 1996).

 Because the correlation function is strong on small scales, the
approximation of isotropy is a good one on these scales, and the
effects of peculiar velocities are large, the measurement of this
small-scale pairwise velocity dispersion is quite straightforward, and
indeed, measurements of this quantity have been done from the
anisotropy of the correlation function for many surveys (cf., Davis \&
Peebles 1983b; Mo, Jing \& B\"orner 1993; Fisher \etal\ 1994b; Marzke
\etal\ 1995; Guzzo \etal\ 1995; Somerville, Davis, \& Primack 1996).
However, the results these workers have found have ranged over a large
factor.  The problem is that
this statistic, as its name implies, is pair-weighted, which means
that the densest regions (the rarest, richest clusters) where the
velocity dispersion is the highest, tend to dominate the
measurement. Therefore, a given measurement of the velocity dispersion
is very sensitive to the presence or absence of the richest clusters
in the survey volume. There have been several attempts to invent
variants on the statistic that are less sensitive to this problem, and
thus measure the velocity dispersion in the field. The point is that
there is a fair amount of evidence from observations of the peculiar
velocity field of galaxies (cf., the contribution by Willick to this
volume) that indicates that the velocity field outside of clusters is
very quiet (e.g., Sandage 1986; Brown \& Peebles 1987; Groth,
Juszkiewicz, \& Ostriker 1989; Burstein 1990; Ostriker \& Suto 1990;
Strauss, Cen \& Ostriker 1993), and that a measure of this should be a
sharper discriminator of models than the classic velocity dispersion
measure.

\subsubsection{Correcting the Density Field for Peculiar Velocities}
\label{sec:iteration}

\begin{figure}
\fig 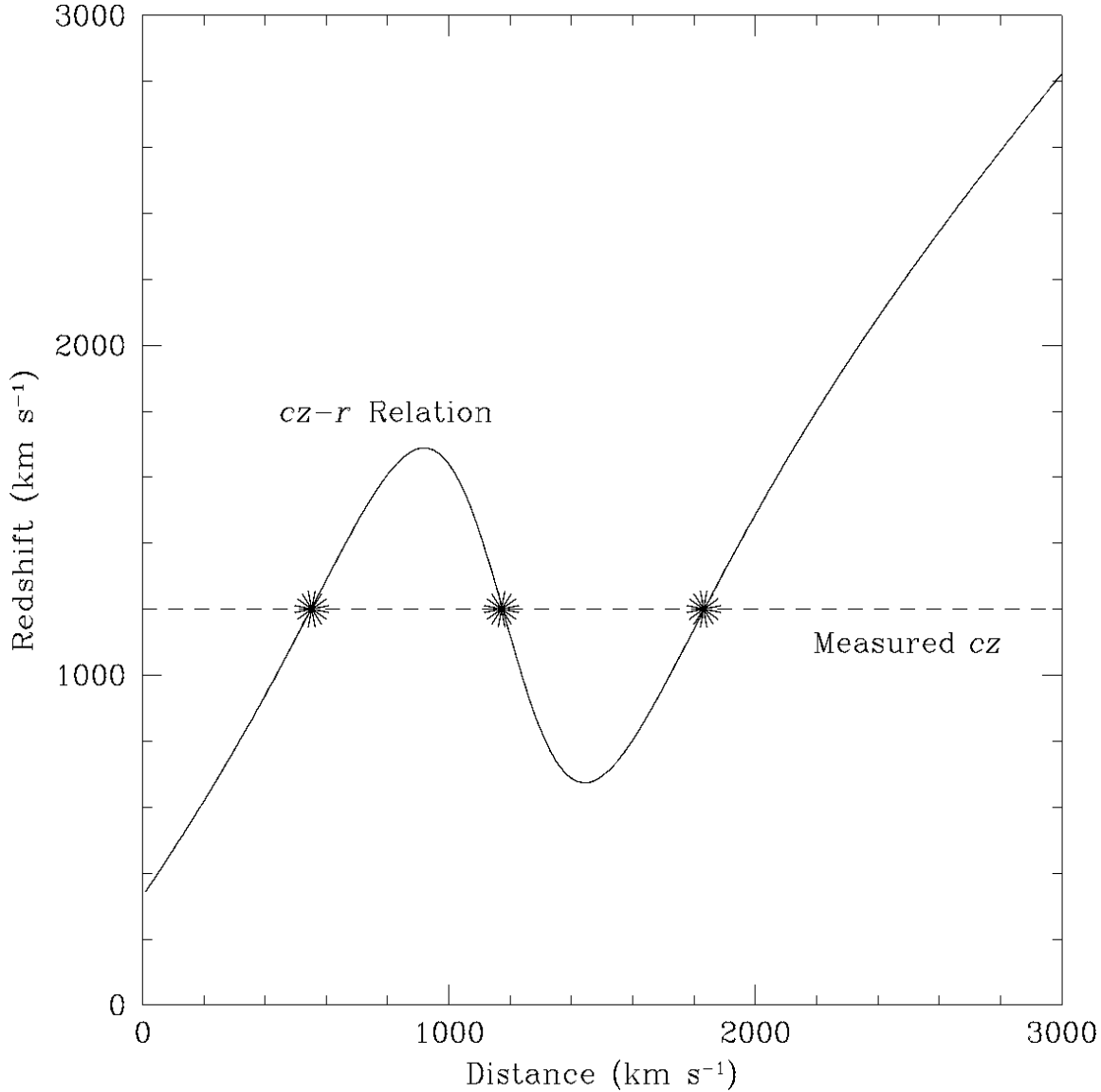, 6, 6
\caption{The redshift-distance diagram in the vicinity of a cluster.
Infall into the cluster causes a region in which the redshift does not
climb monotonically with distance.  A galaxy with redshift 1200 \kms\
in this direction could therefore be at three distinct distances, as
indicated by the stars.}
\label{fig:tvz}
\end{figure}

We have discussed the relationship between the clustering of galaxies
in real space and redshift space, and found that statistical measures
of clustering such as the power spectrum differ systematically in the
two cases.  However, there are situations in which we want to do more
than simply ask for corrections to statistical quantities measured in
redshift space.  In particular, it is quite straightforward to measure
the density field $\delta(\bfs)$ of galaxies in redshift
space\footnote{\bfs\ is the standard notation to refer to redshift
space, in contradistinction to the real space \bfr.}, but how might we
correct the resulting map for peculiar velocities?  In linear theory,
we have an answer to this question, due to the direct relation between
the density and peculiar velocity field, in the form of
Eq.~(\ref{eq:linear-theory}) or its integral equivalent:
\begin{equation} 
\bfv(\bfr) = {\beta \over {4\,\pi}}\int\! { \delta(\bfr')
(\bfr' - \bfr) \d^3 \bfr' \over \left|\bfr' - \bfr\right|^3}\quad,
\label{eq:v(delta)} 
\end{equation}
where we have assumed linear biasing as before.  This equation offers
an approach to correcting the distribution of galaxies explicitly for
peculiar velocities if one has a moderately deep, full-sky redshift
survey, at least on scales large enough that linear theory is likely
to hold (Yahil \etal\ 1991).  One measures the quantity $\delta$ of
the galaxies, and solves for the resulting velocity field using
Eq.~(\ref{eq:v(delta)}).  One then corrects the redshifts of each
galaxy accordingly\footnote{Keep in mind that $H_0 \equiv 1$ with our
units.}:
\begin{equation} 
r =  cz - \hat \bfr \cdot \bfv(\bfr)\quad,
\label{eq:cz-correct} 
\end{equation}
where $\hat \bfr$ is the unit vector in the direction of the galaxy in
question, and $\bfv$ is the peculiar velocity predicted by
Eq.~(\ref{eq:v(delta)}) at that position.  Of course, this changes the
position of each galaxy, and therefore the density field, and thus
this process must be done iteratively until convergence.  There are a
number of points to be made here:
\begin{itemize}
\item The corrections clearly depend on an assumed value of $\beta$,
and this analysis in and of itself yields no estimate of $\beta$.
Therefore, in practice, one produces a separate density field solution
for each value of $\beta$ one might be interested in. 
\item Eq.~(\ref{eq:v(delta)}) is only valid on linear scales, and
therefore one must in practice smooth the density field on small
scales in applying this equation.  A typical smoothing that is used is
a Gaussian with $\sigma = 500 \kms$. A related problem is that
clusters of galaxies typically have quite large velocity dispersions,
which are very far from being describable by linear theory.  In
practice, one collapses the galaxies associated with the prominent
clusters to a single redshift, to suppress this behavior. 
\item In a flux-limited redshift survey, the shot noise in the density
field necessarily increases as a function of distance from the Local
Group.  Therefore one needs to carry out some sort of adaptive
smoothing as a function of distance to suppress the shot noise.  One
possibility is to simply have a smoothing length that increases as the
mean intergalaxy separation.  Another is to expand the density field
not in Cartesian coordinates, but rather in spherical harmonics (cf.,
Nusser \& Davis 1994; Fisher \etal\ 1994c, 1995b).  A third possibility is
to filter the density field with a Wiener or related filter.  As
described in Hoffman's contribution to this volume, these filters are
optimal in the sense of suppressing the shot noise while giving the
minimum variance difference between the derived and true density
fields.
\item In principle, the integral in Eq.~(\ref{eq:v(delta)}) extends to
infinity, while redshift surveys are clearly finite.  This is not as
serious a problem as it may seem.  Redshifts are measured in
the rest frame of the Earth, and are easily corrected to the
heliocentric frame, and to the rest frame of the barycenter of the
Local Group (cf., Yahil, Tammann, \& Sandage 1977).  We of course know
that the Local Group is moving with respect to the rest frame of the
CMB at $\approx 620 \kms$ (e.g., Kogut \etal\ 1993), but we can use
a full-sky redshift survey and Eq.~(\ref{eq:v(delta)}) to {\it
predict\/} this motion (e.g., Strauss \etal\ 1992b; SW).  With this
predicted peculiar velocity at $\bfr = 0$, Eq.~(\ref{eq:cz-correct})
is modified to:
\begin{equation} 
r = cz_{\rm Local\ Group} - \hat \bfr \cdot \left(\bfv (\bfr) - \bfv({\bf 0})\right)\quad.
\label{eq:cz-correct-lg} 
\end{equation}
Here it is apparent that any bulk flow induced by density fluctuations
outside the volume surveyed cancel out; all that is relevant are
higher-order (and therefore intrinsically weaker) multipoles of the
large-scale velocity field.
\item The redshift-distance relation along any line of sight (as given
by Eq.~\ref{eq:cz-correct-lg}) is not necessarily monotonic.  In
particular, in the vicinity of clusters of galaxies, one can get {\it
triple-valued zones}, in which a single redshift can correspond to
three distinct distances.  This is illustrated in Fig.~\ref{fig:tvz},
which shows the relation between the redshift and distance
(Eq.~\ref{eq:cz-correct-lg}) along a line of sight that passes close
to a large mass concentration.  Infall into the cluster causes the
redshift to be greater than the distance on the near side of the
cluster, and further than the distance on the far side.  Thus a galaxy
at $cz = 1200 \kms$ can be at three distinct distances, as indicated
by stars.  Methods for dealing with this are described in Yahil \etal\
(1991) and SW. 
\item There are other methods than the iterative method outlined here
for correcting the density field for peculiar velocities.  In
particular, Nusser \& Davis (1994) and Fisher \etal\ (1995b) describe
non-iterative methods which involve expanding the density field in
spherical harmonics and correcting each mode individually for
peculiar velocities.  The latter authors compare various different
methods by means of of $N$-body simulations, and conclude that all do
roughly equally well. 
\end{itemize}

  Although these techniques indeed do correct the galaxy density field
for redshift distortions, they simultaneously make a prediction for
the velocity field at every point in space.  This is very interesting,
for it allows a direct comparison with the observed peculiar velocity
field, which allows gravitational instability theory (in the form of
Eq.~\ref{eq:linear-theory} or \ref{eq:v(delta)}) to be tested, and to
constrain the value of $\beta$.  This is reviewed in detail in Dekel's
contribution to this volume (cf., Dekel 1994; SW). 

\section{The Relative Distribution of Galaxies and Dark Matter}
\label{sec:bias}

In this review, I have concentrated on the power spectrum as a measure
of the clustering of galaxies. There is quite a variety of other
useful statistics with which to measure the clustering properties of
galaxies, as reviewed in SW and Borgani (1995). The power spectrum is
a complete statistical description of the distribution of galaxies to
the extent that the phases of the Fourier modes are randomly
distributed, which, as we saw above, implies that the one-point
distribution function of densities smoothed on a given scale is
Gaussian.  This may not hold on large scales due to non-Gaussian
initial conditions in the density field (cf., the discussion of
Weinberg \& Cole 1992) or the generation of non-Gaussian terms by
gravitational instability (cf., the review by Juszkiewicz \& Bouchet
1996). On small scales, the distribution is manifestly non-Gaussian
(as it must be on scales on which $\sigma^2 \rightarrow 1$ (cf.,
Eq.~\ref{eq:sigma-Pk}) because $\delta \ge -1$ by definition).
Higher-order versions of the power-spectrum and correlation function
statistics, counts-in-cell statistics, topological statistics,
multi-fractals, and many others have been invented to measure these;
again, see the reviews cited above.  

As mentioned above, one of the big uncertainties in using the
observed distribution of galaxies to infer cosmological quantities is
the unknown relative distribution of galaxies and the dark matter. As
Dekel discusses in this volume, Eq.~(\ref{eq:linear-theory})
means that measurements of the galaxy velocity field yields a measure
of the distribution of {\it all\/} gravitating matter, not just that
component which is apparent in galaxies.  To the extent that linear
biasing holds (Eq.~\ref{eq:linear-bias}), this means that
density-velocity comparisons in the linear regime are able to
constrain the quantity $\beta \equiv \Omega^{0.6}/b$, and not $\Omega$
and $b$ separately.  If one goes beyond linear perturbation theory,
and can successfully {\it measure\/} second-order effects in the
velocity-density relation, then this degeneracy can be broken in
principle.  However, second-order effects are important when the
density contrast becomes comparable to unity. On the smoothing scales
where this is true, there will be regions of space where $\delta_{\rm
dark\ matter}$ approaches $-1$, that is, true voids.  It is clear that
when this happens, the linear biasing model,
Eq.~(\ref{eq:linear-bias}), must break down (at least for $b > 1$, as
is usually assumed); it would predict $\delta_{\rm galaxies} < -1$, a
physical impossibility.  Therefore, if we are to allow ourselves the
generalization to a non-linear relation between velocity and density,
we must also allow ourselves a non-linear relation between dark matter
and galaxy density contrasts, with the resultant increase in the
number of parameters, and the degeneracy between $\Omega$ and the
biasing parameter(s) remains.  The results quoted above in
\S~\ref{sec:Pk-results} should be kept in mind, however, when
considering these complications: to the extent that the biasing
process is {\it local}, that is, that the process of galaxy formation
is a function of the physical properties of matter within a few Mpc of
the galaxy in question, the bias parameter, as measured by the ratio
of the power spectra of the galaxies and the dark matter, appears to
be independent of scale (Weinberg 1995; Kauffmann \etal\ 1996).

Another interesting handle on bias comes through measurements of
higher-order correlation functions; their scaling with scale, it turns
out, depends non-trivially on the bias factor(s) (Fry \& Gazta\~naga
1993; Juszkiewicz \etal\ 1995; Fry 1994; Gazta\~naga 1995; Mo, Jing,
\& White 1996).  The observed amplitudes of the high-order
correlations as a function of scale are sufficiently close to those
predicted from gravitational instability theory without the
complications of bias as to make several of the above papers argue for
the bias being quite small (i.e., $|b - 1| \approx 0$).

  There is a more indirect approach to the bias question, however.
Although we scarcely understand the process of galaxy formation in any
detail (cf., the contribution by White in this volume), we imagine
that there existed astrophysical processes in the early universe which
caused the relative distribution of galaxies and dark matter to differ
from one another.  If this is the case, we would expect generically
that because the astrophysics of formation of galaxies of different
types should be different, these various galaxies should be
distributed differently with respect to each other.  That is, there
should exist a {\it relative\/} bias of galaxies of different types.
This is of course straightforward to measure from redshift surveys. 

  There is one such bias which has been known about since the time of
Hubble: although elliptical galaxies make up only 10-20\% of the
galaxies in the field, they dominate completely in the cores of
clusters of galaxies.  That is, the density fields of elliptical and
spiral galaxies are quite different in the densest regions (cf.,
Dressler 1980, 1984; Postman \& Geller 1984; Whitmore \etal\
1993), something that is quite apparent in redshift space maps of
spirals and ellipticals separately showing the relative distribution
of each (e.g., Giovanelli, Haynes \& Chincarini 1986; Huchra \etal\
1990). It remains quite unclear how this relative bias continues into
the field.

  There are two ways one might imagine measuring the relative bias of
two samples of objects.  The first is to compare them statistically:
compute the power spectrum, the correlation function or higher-order
statistics for each, and determine the bias accordingly. Thus, for
linear biasing, the ratio of the two power spectra is proportional to
the relative bias parameter squared, while the $S_3$ parameter
(defined as $\vev{\delta^3}/\vev{\delta^2}^2$ smoothed on some scale)
should scale as $1/b$.  SW give a brief review of the extensive
literature on determinations of relative bias between samples using
this approach.  However, the statement
$P(k)_{\rm galaxies} = b^2 P(k)_{\rm dark\ matter}$ is clearly a much
weaker statement than is Eq.~(\ref{eq:linear-bias}) (for the latter to
hold, one not only needs a specific relation between the amplitudes of
the Fourier modes of $\delta$, but also that the phases of the modes
agree), and if two samples are contained in the same volume, it is a
much more powerful tool simply to compare the density fields of the
two samples directly.  There is much work that has gone in this
direction in the guise of looking for a population of objects that
``fills the voids'' defined by the distribution of ordinary galaxies.
One elegant statistical approach to quantifying relative bias which
takes the phase information into account is the cross-correlation
statistic: one asks for the mean number of galaxies of sample 2 in
excess of random a given distance from galaxies of sample 1.  Another
approach is to take Eq.~(\ref{eq:linear-bias}) literally, and compare
the density fields of the two samples point by point (cf., Strauss
\etal\ 1992a; Santiago \& Strauss 1992).

  Despite a great deal of work on this question using a large variety
of samples, the results can be summarized in a few sentences. On
scales larger than 5\mpc\ or so, there is no direct evidence for
relative biasing that is non-linear or a function of scale, although
the limits on such effects are rather weak\footnote{On small scales,
at least, the slopes of the two-point correlation functions of
ellipticals and spirals are different (Davis \& Geller 1976;
Giovanelli \etal\ 1986).}.  Late-type galaxies show a
weaker clustering signal than do early type galaxies, by a factor of
1.5 to 2.0 in the two-point correlation statistics.  This manifests
itself in a number of ways: one can find papers in the literature that
show that clustering strength increases with the luminosity, central
surface brightness, redness, radio power, and mass of the population
of galaxies, and decreases with their infrared emission and
emission-line strength.  Of course, all these quantities are
correlated with morphological type, and so it remains unclear whether
these trends are separate from the correlation of clustering strength
with Hubble type.  Moreover, there has yet to be a sample which has
been shown clearly to fill the voids which are so prominent in the
galaxy distribution. 

  If clustering is indeed a function of galaxy luminosity, the
assumption of the universal luminosity function which went into the
definition and use of the selection function (\S~\ref{sec:phi(r)}) is
clearly not valid.  For this reason, as the data get better and the
precision with which we measure statistics such as the power spectrum
improve, we will either have to include elaborate models for the
dependence of clustering strength on luminosity in our analyses, or we
will have to do analyses on subsamples of our data over narrow slices
in luminosity.

\section{Surveys for the Future}
\label{sec:future}

 In reviewing this field, I am struck by the progress that has been
made over the last decade or so.  We are now measuring, or at least
constraining, quantities such as the large-scale bias, the power
spectrum on the largest scales, and the value of $\Omega$, which we
had very little hope of getting a handle on at the time people started
carrying out large-scale redshift surveys at the end of the 1970's.
Part of this progress has been a realization, as is so common in
astrophysics, that as we learn more about the systems we study, we
realize how na\"\i ve and over-simplified our analyses and assumptions
have been, and how much more complicated (and interesting!) reality
is.  In the early 1980's, for example, people were modelling infall
into the Virgo cluster (cf., Davis \& Peebles 1983a) as if it were an
isolated overdensity in a uniformly distributed sea of galaxies. A
glance at Figs.~\ref{fig:sgp} and \ref{fig:pie} tells us that this is
far from a valid model\footnote{although Virgocentric infall remains a
useful cosmological probe; see the contribution by Peebles in this
proceedings.}!

  That having been said, we are really only starting to be able to
measure cosmological quantities of interest. For example, the values
of $\beta$ determined from redshift space distortions, analyses of the
peculiar velocity field (cf., Dekel, this volume; Shaya,
Peebles \& Tully 1995), and the mass-to-light ratio of virialized
systems (e.g., Bahcall, Lubin \& Dorman 1995) vary from 0.3 to over
unity, with little convergence between methods in sight at the moment
(cf., Fig.~20 of SW).  The amplitude of the galaxy power spectrum on
large scales is uncertain by at least a factor of two, and probably
more, and the relative biasing of galaxies of different types
(\S~\ref{sec:bias}) is only partially to blame for the ambiguity. In
other words, we are only just starting to do real quantitative
cosmology with large-scale surveys, and I imagine that there is much
exciting, and unanticipated, science waiting to be discovered when we
start measuring quantities to 10\% accuracy (cf., the flurry of
theoretical and observational activity prompted by the recent
convergence of measurements of $H_0$ to within 25\%, as summarized in
the contributions of Tammann and Freedman to Turok 1996).  To beat
down the errors further requires much more massive redshift surveys,
with:
\begin{itemize}
\item Much larger volume surveyed.
\item Much larger number of galaxies with redshifts.
\item Much tighter control of statistical and especially systematic
errors of the photometric quantities by which galaxies are selected.
\end{itemize}

Here I discuss two such surveys in the advanced planning stages: the
Sloan Digital Sky Survey (SDSS) and the Two-Degree Field Survey
(2dF).  As I am involved with the former, and thus much more familiar
with it, I will put the greater emphasis here on it. 

\subsection{The Sloan Digital Sky Survey}
\label{sec:SDSS}

The SDSS is a collaboration between Princeton University, the
Institute for Advanced Study, the University of Chicago, the Fermi
National Accelerator Laboratory, the University of Washington, Johns
Hopkins University, the United States Naval Observatory, and the Japan
Promotion Group.  We are building a dedicated 2.5m large-field optical
telescope at Apache Point in South-Eastern New Mexico, which should
see first light late this fall (1996).  The telescope will have two
main instruments: a photometric camera with 30 $2048\times 2048$, and
24 $2048 \times 400$ SIte CCD chips on its focal plane, and a pair of
double multi-object spectrographs, together taking 640 fibers of $3''$
aperture.  The purpose of this survey is several-fold.  With the
photometric camera, the one-quarter of the sky (roughly 10,000 square
degrees) centered on the Northern Galactic Cap will be surveyed in
drift-scan mode through five broad-band filters ($u', g', r', i', z'$,
a new photometric system (cf., Fukugita \etal\ 1996) with effective
wavelengths of 3540\AA, 4760\AA, 6280\AA, 7690\AA, and 9250\AA,
respectively) almost simultaneously, with an effective exposure time
of 55 sec for each.  Stellar objects will be detected at $5 \sigma$ to
$r' \approx 23$.  From the resulting list of objects detected in the
photometric survey, galaxies will be selected to a photometric limit
of $r' \approx 18$, and quasar candidates will be selected from the
stellar objects by their distinctive colors to a magnitude and a half
fainter. These objects, with a density of roughly 120 per square
degree, will be observed with the multi-object spectrograph to obtain
spectra from 3900-9100\AA\ with a resolution of 2000.  Over the course
of the survey, which is expected to take five years, we will thus
obtain spectra of roughly $10^6$ galaxies and $1.5 \times 10^5$ quasar
candidates, of which we hope 60-70\% will be {\it bona fide\/}
quasars.  Finally, during the Fall months, when the Northern Galactic
Cap is unreachable, we plan to repeatedly scan an equatorial stripe
$2.5^\circ$ wide and $90^\circ$ long in the South Galactic Cap,
centered on $\delta = 0^\circ$. This will allow us to do photometry
over 225 square degrees to a photometric limit roughly two magnitudes
fainter than in the North.  Further photometry (only a single pass)
will be done on two ``outrigger'' great circle stripes centered
roughly at $\delta = +15^\circ$ and $\delta = -10^\circ$ (cf.,
Fig.~\ref{fig:surveys}) to maximize the number of baselines for
measurements of the power spectrum on the very largest scales.

 This is a very quick overview of the plans for the SDSS. Reviews can
be found by Gunn \& Knapp (1993) and Gunn \& Weinberg (1995), and a
great deal of technical detail can be found in the text of our
proposal to the NSF at {\tt http://www.astro.princeton.edu/GBOOK/} on
the World-Wide Web.  Here let us summarize the goals and prospects for
large-scale structure studies with the data from the SDSS.  We have
space here to touch upon only a few scientific issues which the SDSS
will impact; for many more (both in large-scale structure, and in many
other fields of extragalactic and Galactic astronomy), see the Web
site above. 

  The SDSS redshift survey will survey to a depth that has been probed
previously; the depth is comparable to that of the LCRS and ESP
surveys discussed above in \S~\ref{sec:variety}.  Moreover, like the
LCRS, the survey galaxies will be selected from photometric CCD data.
However, the the volume covered by the SDSS redshift survey will be
enormously greater; the solid angle coverage of the SDSS will be 14
times that of the LCRS.

\begin{figure}
\fig 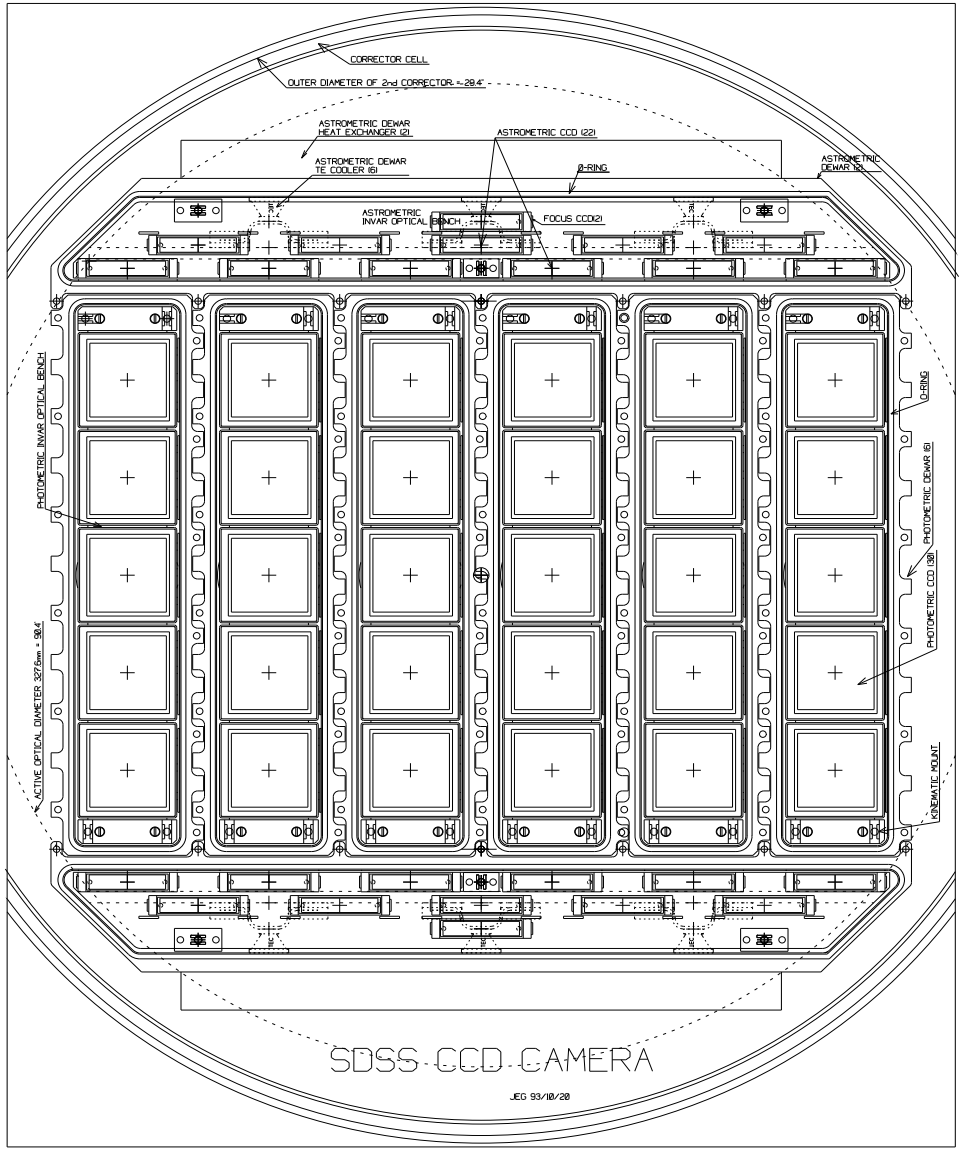, 6, 5
\caption{A schematic of the focal plane of the photometric camera of
the SDSS. The camera works in drift scan mode, scanning in the
vertical direction, 
and thus traces out six parallel strips on the sky.  The radius of the
focal plane (as indicated by the dashed circle) is roughly 2.5 degrees
in diameter. Each of the large CCD's is $2048\times 2048$, with $0.4''$
pixel size.  Each of the five rows of CCD's has a different
photometric filter in front of it, so that a given area of sky is
scanned in $r', i', u', z',$ and $g'$, in order, over a span of
roughly 8 minutes.  There are 22 $2048\times 400$ CCD's with $r'$
filters and 4.2 mag neutral density filters, at the leading and
trailing edges of the arrays. These chips will saturate at $r' = 6.6$
(rather than $r' \approx 14$ for the large chips), allowing stars in
astrometric catalogs (especially HIPPARCOS, cf., Kovalevsky \etal\
1995) to be tied to the SDSS images.  Finally, the two $2048 \times
400$ CCD's at the top and bottom of the array are used to measure and
adjust the focus.}
\label{fig:sdss-camera}
\end{figure}

Large-scale structure studies are a major goal of the SDSS.  Thus a
great deal of emphasis has been put on controlling systematic errors
in the survey.  The focal plane of the photometric camera is shown in
Fig.~\ref{fig:sdss-camera}.  The camera will draft-scan at sidereal
rate along great circles in the sky and thus observe six parallel
strips of the sky simultaneously; two passes separated by roughly
$13'$ will cover a stripe $2.5^\circ$ wide, with roughly $2'$ overlap
between strips.  Astrometric calibration is done with the astrometric
chips, as described in the caption to Fig.~\ref{fig:sdss-camera}.
Photometric calibration will be done with a separate dedicated 24-inch
telescope at the Apache Point site, which will spend the night
obtaining photometry of a series of roughly 30 primary standards
through the SDSS filters to measure the extinction and photometricity
of the sky every hour, and then tie this solution to the photometry of
the 2.5m telescope by observing secondary patches in regions of the
sky covered by the stripes.  The photometric calibration will be
checked {\it a posteriori\/} with the overlaps between survey stripes,
and probably also with a series of stripes taken perpendicular to the
main great circle scans of the sky.

  We of course want a sample of galaxies with accurate photometry {\it
as they would be seen from outside our Galaxy}.  Therefore we plan to
measure Galactic extinction using our multi-color data. In particular,
we will use the colors of distant hot halo subdwarfs (selected by
their colors, and confirmed spectroscopically), as well as number
counts and color distributions of the faint galaxy populations we
find, possibly supplemented by H$\rm\scriptstyle I$ maps from Stark
\etal\ (1992) and Burton \& Hartmann (1994), and the long-wavelength DIRBE maps
from Schlegel, Finkbeiner \& Davis (cf., Schlegel 1995). 

  Galaxies will be selected for spectroscopy from the photometrically
calibrated images after an {\it a priori\/} correction for reddening
either from Burstein \& Heiles (1982) or Schlegel
(1995)\footnote{Given the subtleties of the determination of reddening
using the methods mentioned above, we will be unable to create an
improved reddening map ``on the fly'', thus our final large-scale
structure analyses will require correction for the difference between
our {\it a priori} and final reddening maps.}.  We wish to select
galaxies in as uniform a way as possible, minimizing the effects of
redshift.  Ideal would be to measure ``total'' fluxes for galaxies,
but because galaxies are extended objects, without sharp edges, any
``total'' flux one actually measures for them is either a function of
the sky level and the depth of the image, or requires a
model-dependent extrapolation.  Isophotal fluxes have the drawback of
being affected by extinction and cosmological surface-brightness
dimming in complicated ways, as well as being ill-defined for low
surface brightness galaxies.  We have thus opted for selecting our
galaxy sample based on {\it Petrosian\/} (1976) fluxes in the $r'$
band\footnote{We have looked into the possibility of a joint selection
in $r'$ and one of the bluer bands. However, the larger extinction in
the atmosphere and the lowered throughput at $u'$ makes this band
impractical, and simulations have shown that the galaxy populations
selected in $g'$ would be almost indistinguishable from those at $r'$,
both in redshift distribution and morphological mix.}. Let $I(\theta)$
be the azimuthally averaged surface brightness profile of a galaxy in
the $r'$ band.  We define the Petrosian {\it ratio\/} ${\cal R}_P$ as
the ratio between the local and integrated surface brightness profile
at radius $\theta$; in practice:
\begin{equation} 
{\cal R}_P(\theta) = {{\int_{0.8 \theta}^{1.25 \theta} I(\theta')\, 2 \pi\,
\theta'\, \d \theta' /[\pi\, (1.25^2 - 0.8^2) 
\theta^2]} \over
{\int_{0}^{\theta} I(\theta')\, 2 \pi\, \theta'\, \d \theta' /[\pi\,
\theta^2]}}\quad.
\label{eq:Petrosian-ratio} 
\end{equation}
We then define the Petrosian {\it radius\/} $\theta_P$ as that radius at which
the Petrosian ratio falls to some specified level, say 1/8:
\begin{equation} 
{\cal R}_P(\theta_P) = 0.125\quad.
\label{eq:Petrosian-radius} 
\end{equation}
The Petrosian {\it flux\/} $f_P$ is then that measured within a fixed
number (say 2) of Petrosian radii:
\begin{equation} 
f_P = \int_0^{2 \theta_P} I(\theta')\, 2 \pi\, \theta'\, \d\theta'\quad.
\label{eq:Petrosian-flux} 
\end{equation}

It is this latter quantity (suitably turned into magnitudes) on which
we will select our galaxies (with perhaps some further adjustments in
the values of the constants 0.125 and 2 in
Eqs.~\ref{eq:Petrosian-radius} and \ref{eq:Petrosian-flux}) . Because
the definition of Petrosian quantities is based on the surface
brightness profile of the galaxy itself, the Petrosian radius will be
the same metric radius on a galaxy, independent of its distance to us
or foreground extinction, and tests have shown that at least to our
spectroscopic limit, it depends only very weakly on effects of seeing
or noise.  It also has the advantage of being definable for galaxies
of all types, including those of very low surface
brightness\footnote{One important exception to this last statement is
galaxies with power-law surface brightness profiles, as many cD
galaxies are (e.g., Postman \& Lauer 1995, and references therein);
for such galaxies, the Petrosian ratio asymptotes to a constant that
can be above the level at which the Petrosian radius is defined. In
such cases, we will switch over to an aperture magnitude. Another
potential problem is galaxies with hierarchical structure (e.g., a
Seyfert nucleus in a disk galaxy) for which the Petrosian ratio may
not be monotonic.  Given our chosen value of the value of Petrosian
ratio at which the Petrosian radius is defined, this is a problem only
for a very small fraction of galaxies.}.  This last point has a
drawback; we cannot obtain spectra of galaxies of arbitrarily faint
central surface brightness, no matter how bright their Petrosian
fluxes are.  We could avoid this problem by cutting on the amount of
light entering the $3''$ aperture of the fibers.  However, such a cut
would be very difficult to model as a function of redshift.
Therefore, we instead include a secondary cut on Petrosian surface
brightness at $\mu_P \le 22$ in $r'$.  This quantity is defined as
follows: we define a Petrosian half-light radius $\theta_{50}$, such
that:
\begin{equation} 
\int_0^{\theta_{50}} I(\theta')\, 2 \pi\, \theta'\, \d\theta' = 0.5 f_P\quad;
\label{eq:Petrosian-halflight} 
\end{equation}
the Petrosian surface brightness is then:
\begin{equation} 
\mu_P = {{0.5 f_P} \over \pi\,\theta_P^2}\quad.
\label{eq:Petrosian-SB} 
\end{equation}
   Simulations show that the
resulting galaxy sample has a distribution of $3''$ aperture
magnitudes with a sharp cut-off at $r' = 19.5$, at which point the
signal-to-noise ratio is still adequate to obtain redshifts for our
exposure time ($\sim 45$ min).  It is unfortunate that the limitations
of exposure time and the size of our aperture do not allow us to
obtain redshifts for the very low surface brightness population of our
galaxies, which show an intriguingly different large-scale
distribution from those of ``ordinary'' galaxies (cf., Mo, McGaugh, \&
Bothun 1994).

There is another class of galaxies to be targeted spectroscopically in
the SDSS.  Luminous red elliptical galaxies tend to be metal-rich, and
thus have strong and prominent metal absorption lines.  This means
that redshifts can be measured for them at a lower signal-to-noise
ratio than for typical absorption-line galaxies.  These objects are
interesting in their own right, because they are often associated with
clusters; indeed, brightest cluster galaxies make up the most luminous
and red population of galaxies, and thus targeting the luminous red
ellipticals allows us to obtain redshifts for a deep sample of
clusters.  We will estimate redshifts photometrically from the
five-color data for all galaxies using the methods of Connolly \etal\
(1995) (which should be especially accurate for quiescent red
ellipticals), and thereby determine a luminosity.  Cuts will be made
in luminosity and K-corrected color (the most luminous elliptical
galaxies are very uniform in color, which is why the photometric
redshift determination is so accurate for this class of objects),
yielding a spectroscopic sample of $\sim 10^5$ objects volume-limited
roughly to a redshift of $0.45$.

 I have gone into quite a bit of detail about the selection criteria
for our galaxies in order to emphasize the care that is being taken to
obtain a uniform sample, as free as possible of biases and systematic
errors.  The scientific goals of the project range from large-scale
structure studies, which are the subject of this review, through
quasars and the global properties of galaxies, to Galactic structure
and interstellar extinction.  Again, see our NSF proposal at the URL
quoted above for more detail. 

\begin{figure}
\fig 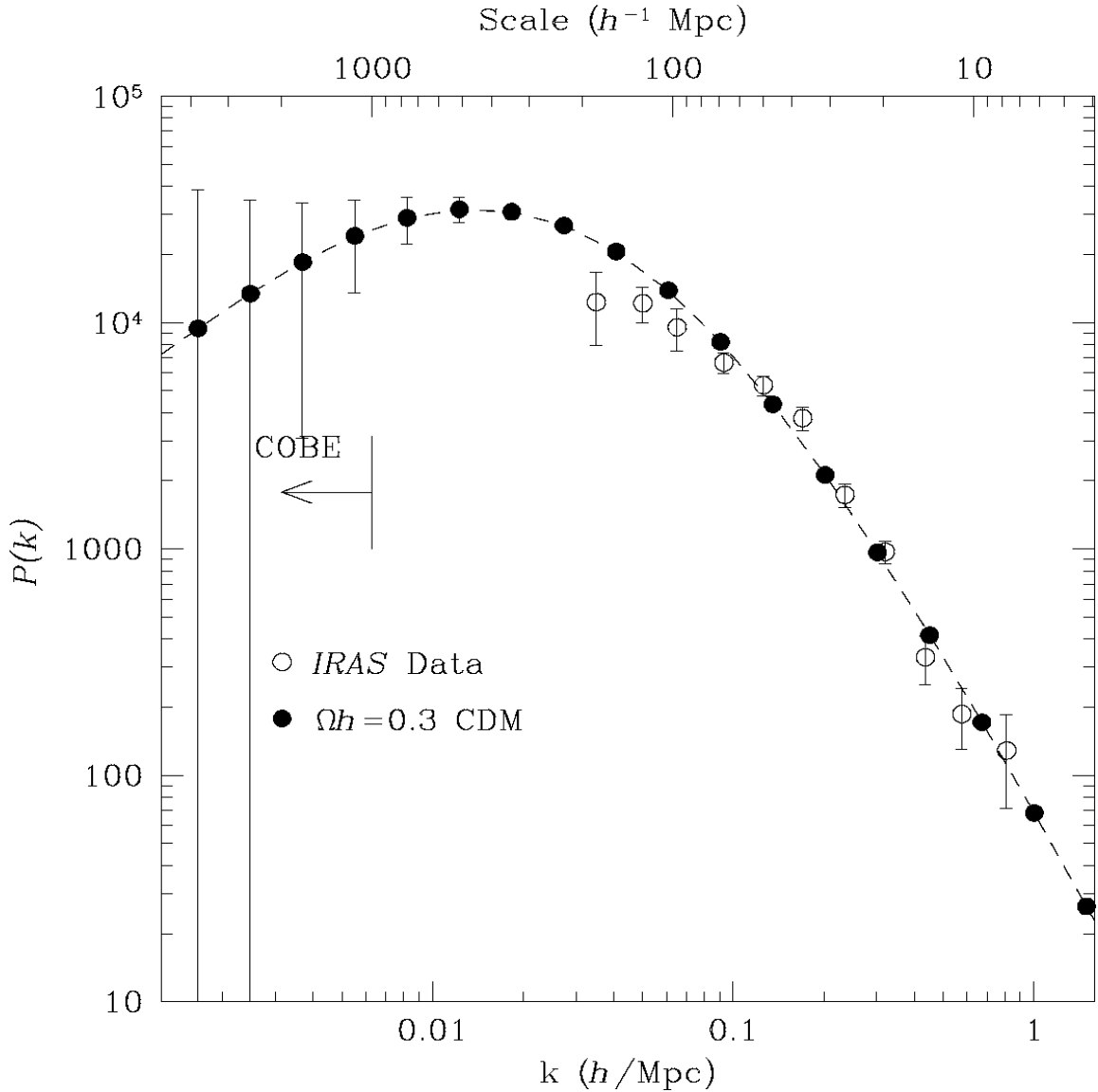, 6, 6
\caption{The observed power spectrum of \iras\ galaxies from Fisher
\etal\ (1993) (open circles) is shown on top of the linear theory
$\Gamma \equiv \Omega h = 0.3$ CDM power spectrum, with error bars at
selected points expected for the SDSS volume, using the formulae of
Feldman \etal\ (1994).  The physical scale corresponding to the
$10^\circ$ resolution of the COBE satellite is indicated.}
\label{fig:sdss-power}
\end{figure}

  The large-scale structure goals of the survey are manifold.  The
most obvious is of course the measurement of the power spectrum on
very large scales. Fig.~\ref{fig:sdss-power} shows the measured power
spectrum of \iras\ galaxies from Fisher \etal\ (1993), together with a
prediction of what the SDSS will see ($\Gamma = 0.3$ CDM), with error
bars following the formalism of Feldman \etal\ (1994).  Thus the SDSS
will measure the galaxy power spectrum on scales on which the
Sachs-Wolfe effect is directly measured by COBE.  Indeed, it will
measure $P(k)$ with great precision on smaller scales in the CMB which
are now just starting to be probed by balloon and ground-based
missions, and which will be studied in detail by the next generation
of satellites (cf., Bennett \etal\ 1995; Mandolesi \etal\ 1995).  If
we have sufficient control on systematic errors, the red luminous
elliptical sample described above should be able to probe
$P(k)$ on the larger scales with appreciably smaller error bars than
are shown in Fig.~\ref{fig:sdss-power}. 

  The redshift-space distortions discussed above in
\S~\ref{sec:redshift-distortions} have allowed determinations of
$\beta$ to an accuracy of 50\% or so, and even that has required a
great deal of modelling (cf., the discussion in Fisher \etal\ 1994b).
As we have mentioned above, the principal limitation of these analyses
is the smallness of the volume probed; for this statistic, we are far
from a fair sample\footnote{As this example makes clear, the
definition of a ``fair sample'' of the universe depends very much on
the sort of statistic one is trying to measure.  Thus current samples
are more than large enough to define the mean number density of
luminous galaxies, but are certainly not adequate to measure the power
spectrum on the largest scales.}.  The SDSS will probe a very large
number of superclusters at (presumably) random inclinations, greatly
reducing the statistical errors in this analysis.  Moreover, one can
probe for more subtle effects which could only be set {\it a priori\/}
in the analysis of Fisher \etal\ (1994b), such as the amplitude of the
mean streaming as a function of scale, and the detailed shape of the
pairwise velocity distribution function on small scales.  There is
even the possibility of measuring the second moment of the velocity
distribution function on {\it linear\/} scales (cf., Fisher 1995),
which has the possibility of breaking the degeneracy between $\Omega$
and $b$.

  But perhaps the most exciting large-scale structure studies to be
done with the SDSS will involve measurements of the distribution of
galaxies as a function of the physical properties of the galaxies
(such as morphological type, color, luminosity, strength of emission
or absorption lines, surface brightness, and so on).  As we have
discussed in \S~\ref{sec:bias}, it is now clear that there exists a
relative bias between galaxies of different types, but we as yet
cannot characterize its nature as a function of scale or its
universality.  It is now becoming clear that the simple linear biasing
model of Eq.~(\ref{eq:linear-bias}) cannot hold true in detail, and
that therefore different analyses of large-scale structure in a sense
are sensitive to different moments of the relationship between the
galaxy and dark matter distribution.  The analyses mentioned above, as
well as many others, will be done not only on the full sample of
galaxies in the SDSS, but also on subsamples divided by the physical
properties of galaxies (the sample will be large enough to allow us to
do this!); comparison of the results will give us strong constraints
on the relative bias of different populations of galaxies.

\subsection{The Two Degree Field}
\label{sec:2DF}

The Two Degree Field (2dF) refers to an instrument that is undergoing
final commissioning at this writing, mounted on the Prime Focus of
the Anglo-Australian Telescope. It is a fiber-fed multi-object
spectrograph with 400 fibers and a two-degree diameter field, as its
name implies.  The fiber coupler is doubled, so that one field can be
prepared with a robot arm while a second one is observed, minimizing
overhead.  This is a general purpose observatory instrument which is
to be used for a variety of surveys, but a collaboration of British
and Australian astronomers is planning to use it for a redshift survey
of galaxies selected from the APM galaxy catalog (cf., Maddox \etal\
1990abc); thus, unlike the SDSS, the 2dF survey does not attempt to
create a galaxy catalog from scratch.  The survey will cover 1700
square degrees, and will obtain redshifts of galaxies to $B = 19.7$
\footnote{This is slightly deeper than the SDSS, given typical galaxy
colors of $r' - B \approx -1$ (Frei \& Gunn 1994).} as measured by the
APM. It will consist of three parts, contiguous strips of $75^\circ
\times 12.5^\circ$ and $65^\circ \times 7.5^\circ$ in the Southern and
Northern Galactic Caps, respectively, and 100 random circular fields
of radius $2^\circ$ over the Southern Galactic Cap.  The survey will
contain roughly 250,000 galaxies, and will take roughly 90 nights of
AAT dark time.  The survey geometry is chosen to maximize sensitivity
to large-scale structure; indeed, the 100 random fields give a variety
of baselines to probe the power spectrum on the largest scales.  The
science goals of the 2dF survey are similar to those of the
large-scale structure goals of the SDSS, although the two surveys
differ on the approach to galaxy catalogs and target selection, the
sky coverage, the depth of the survey, and other details.

  There will be an extension to the 2dF survey done to measure
redshifts for $\sim 6000$ galaxies to $R = 21$ to extend large-scale
structure studies to redshifts of 0.3. The SDSS will be able to do
such studies using the luminous red elliptical sample mentioned above,
as well as from the photometric sample (which of course extends quite
a bit fainter than the spectroscopic sample).  As I am not involved in
the 2dF project, I cannot go into the sort of detail I have for the
SDSS. The 2dF instrument is described in Taylor (1995), a description
of the planned redshift survey can be found at the Web site
\newline{\tt
http://qso.lanl.gov/$\sim$heron/Colless/colless\_heron/colless\_heron.html}.

\section{Conclusions}
\label{sec:conclusions}

\begin{figure}
\centerline{\psfig{file=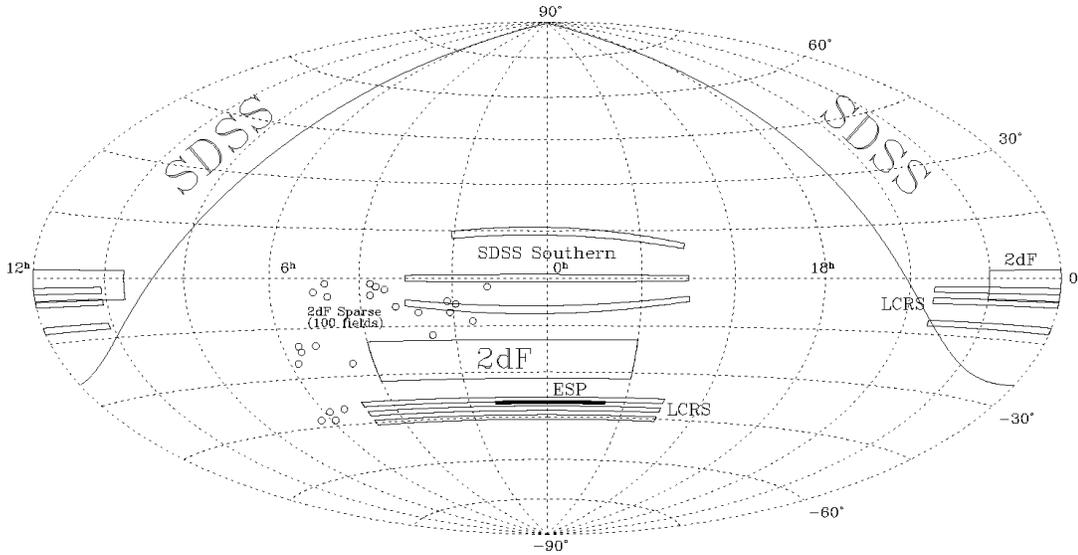,width=7cm,angle=270,bbllx=47.8pt,bblly=270.6pt,bburx=552pt,bbury=627.2pt}}
\caption{An Aitoff projection of the sky in right ascension and
declination, showing the sky coverage of the various surveys discussed
in the text.}
\label{fig:surveys}
\end{figure}

  I have reviewed recent progress in measurements of large-scale
structure with redshift surveys of galaxies in the nearby universe,
with emphasis on the power spectrum of the galaxy density field.
There are a number of complicating factors which separate the ideal
linear power spectrum of the dark matter from the quantity measured,
such as redshift space distortions, non-linear effects, and the
relative biasing of galaxies and dark matter, but in each case, these
complications include in themselves valuable cosmological
information. I conclude with Fig.~\ref{fig:surveys}, modified from a
similar figure in Guzzo (1996), which shows the
coverage on the sky of the various surveys we have discussed. This is
certainly not complete; the \iras\ surveys are not shown here, nor are
the CfA2 and SSRS2 surveys (cf., da Costa \etal\ 1994a) or the
Perseus-Pisces redshift survey (Haynes \& Giovanelli 1989).  However,
all the surveys shown here are probing or will probe the large-scale
distribution of galaxies to redshifts of 0.1 over substantial solid
angles, with more careful and uniform sample selection than has been
possible before now.  We are just starting to measure cosmologically
important parameters with redshift surveys with errors of 50\% or
less; this next generation of surveys hold out great hope to reduce
these errors substantially, with the potential of new and unexpected
insights into our picture about the formation of large-scale structure
and galaxies. 

{\bf Acknowledgements:} I wish to thank my colleagues on the various
redshift surveys I discuss here, as well as my SDSS colleagues whose
work I summarize above.  Avishai Dekel, Michael Vogeley, and David
Weinberg all made useful comments on a previous draft of this paper.
I also thank Huan Lin of the LCRS collaboration for permission to show
Fig.~\ref{fig:pie}, Tsafrir Kolatt for the input files to
Fig.~\ref{fig:kolatt-power}. and Luigi Guzzo for his SM macros which
led to Fig.~\ref{fig:surveys}.  I am happy to acknowledge the
financial support of an Alfred P. Sloan Foundation fellowship.

\end{document}